\documentclass[authoryear,preprint,12pt]{elsarticle}
\usepackage[latin9]{inputenc}
\usepackage{float}
\usepackage{textcomp}
\usepackage{amsmath}
\usepackage{amssymb}
\usepackage{graphicx}
\usepackage{graphicx,caption}
\usepackage{lscape}
\usepackage{array}
\newcommand{\head}[2]{\multicolumn{1}{>{\centering\arraybackslash}p{#1}}{{#2}}}
\PassOptionsToPackage{version=3}{mhchem}
\usepackage{mhchem}
\usepackage{chngpage}

\makeatletter

\DeclareTextSymbolDefault{\textquotedbl}{T1}

\usepackage{adjustbox}
\usepackage{scrextend}

\usepackage{lineno}

\newcommand{\Feratio}{$^{56}\mathrm{Fe}/^{54}\mathrm{Fe}$}
\newcommand{\Mgratio}{$^{25}\mathrm{Mg}/^{24}\mathrm{Mg}$}

\journal{Earth and Planetary Science Letters}

\makeatother

\begin{document}



\title{Isotope velocimetry: Experimental and theoretical demonstration of the potential importance of gas flow for isotope fractionation during evaporation of protoplanetary material }


\author[label1]{Edward D. Young}

\author[label2]{Catherine A. Macris}

\author[label1]{Haolan  Tang}

\author[label2]{Arielle A. Hogan}

\author[label1,label3]{Quinn R. Shollenberger}

\address[label1]{Department of Earth, Planetary, and Space Sciences, UCLA}

\address[label2]{Earth Sciences, IUPUI}

\address[label3]{Lawrence Livermore National Laboratory}

\address{correspondence: eyoung@epss.ucla.edu, camacris@iupui.edu}








\begin{abstract}
We use new experiments and a theoretical analysis of the results to show that  the isotopic fractionation associated with laser-heating aerodynamic levitation experiments is consistent with the velocity of flowing gas as the primary control on the fractionation.  The new Fe and Mg isotope data are well explained where the gas is treated as a low-viscosity fluid that flows around the molten spheres with high Reynolds numbers and minimal drag.  A relationship between the ratio of headwind velocity to thermal velocity and saturation is obtained on the basis of this analysis.  The recognition that it is the ratio of flow velocity to thermal velocity that controls fractionation allows for extrapolation to other environments in which molten rock encounters gas with appreciable headwinds.  In this way, in some circumstances, the degree of isotope fractionation attending evaporation is as much a velocimeter as it is a barometer.  
\end{abstract}
\begin{keyword}
evaporation \sep isotope fractionation \sep pebbles \sep planet formation 
\end{keyword}
\maketitle


\section{Introduction}

Relative volatility is a critical means of partitioning elements and their isotopes in a broad range of astrophysical settings.  In the Solar System there is clear evidence for fractionation by evaporation and condensation.  Evaporation is evidenced unequivocally in some millimeter to centimeter-sized once-molten spherules found in many chondritic meteorites.  These  spherules include igneous calcium-aluminum-rich inclusions (CAIs) and chondrules.  CAIs are rich in refractory elements and among the most ancient of Solar System materials.  Those that have experienced melting exhibit clear evidence for preferential evaporation of light isotopes in the form of enrichment of the heavy isotopes of Mg and Si \citep{Davis1990, Shahar2007}.  Chondrules more closely resemble the compositions of bulk chondrites and generally lack heavy isotope enrichment, despite the fact they too were once molten and floating in space in the early Solar System.  The different isotopic compositions of these objects is attributed to their different environments of melting.  In the case of CAIs, conditions permitted extensive evaporative losses, and thus isotope fractionation, while in the case of chondrules, conditions prevented fractionation.  In particular, CAIs may have evaporated in relative isolation in a near-vacuum while chondrules may have melted in the presence of gas at near equilibrium conditions \citep{Galy2000-1751,Young2004,Cuzzi2006-483}.  At the larger scale, there is a long-standing interest in the effects of evaporation on planetesimals and larger bodies during the magma ocean stage of evolution \citep{Palme1993-979, Kreutzberger1986, Halliday2001, Allegre2001, Paniello2012, Young2019}.

An outstanding issue is the effect that the relative motion between gas and molten bodies has on the evaporation process.  This is arguably critically important to our interpretations of the putative chemical and isotopic effects of evaporation given the prospects for differential motion between condensed phases and gas.  Important examples may include pebble accretion into planetary envelopes \citep[e.g.,][]{Brouwers_2021} and the possibility that chondrules might form in collisional plumes \citep{Lichtenberg2018}.  Indeed, it is possible that by placing so much emphasis on the ambient pressures of the environments where melting occurs, the importance of the relative velocity between gas and melts has gone underappreciated.   Here we try to rectify this oversight.  

In this work we report the results of experiments in which spherules of synthetic enstatite chondrite composition were melted with an infrared (CO$_2$, $10.6\mu \mathrm{m}$) laser while being  levitated in a stream of  N$_2$, and in some instances Ar, gas. We used an enstatite chondrite bulk composition as a first step towards investigating analogs for Earth's progenitors.  Evaporative residues  were analyzed for their Fe and Mg isotopic compositions as well as their Fe and Mg concentrations.  We examined Fe and Mg as representatives of the more volatile major elements and the more refractory major elements, respectively.  Studies of the isotopic effects of evaporation under similar conditions using laser-heating aerodynamic levitation have been reported by \cite{Wimpenny2019}, \cite{Ni2021}, and \cite{Badro2021}.  We show with a mathematical model that these results can be explained by the primary control that the velocity of the gas has on the evaporation process.  The results are potentially important for interpreting the significance of evidence for evaporation in protoplanetary materials.

\section{Methods}

\subsection{Experiments}  
 Vaporization experiments were performed in the High-Temperature Conical Nozzle Levitation (HT-CNL) System (this apparatus is sometimes also referred to as an aerodynamic levitation laser furnace) at the High Pressure and Temperature Geochemistry lab at Indiana University-Purdue University Indianapolis (IUPUI).  The apparatus components and mode of operation  are described in \cite{Ni2021} and further details are provided in Appendix 1. In these experiments, a spherical sample, $\sim 2$ mm in diameter, is levitated by a gas (or gas mixture) issuing from a conical nozzle while being heated with a 40 W CO$_2$ IR laser ($10.6 \mu$m wavelength). The starting material is a synthetic enstatite chondrite powder similar in composition to the EH4 meteorite Indarch made from oxides of the major elements. Heating and levitation take place in a controlled-atmosphere chamber, with gas mixing capabilities. Samples were pre-fused using a broad laser focus at the minimum laser power that affords melting in a stream of Ar gas.  After weighing, pre-fused spherical  samples were heated to 2273 or 2323 K  for 120 to 480 s while levitated in either 99.999\% purity N$_2$, or a mixture of $95\%$ Ar , $\sim 4.5\%$ CO and $\sim 0.5\%$ CO$_2$, the latter mixture imposing an oxygen fugacity relative to the iron-w\"{u}stite equilibrium value, $\log{f_\mathrm{O_2}} -\log{f_\mathrm{O_2}}({\rm IW})$, of $-0.5$ \citep[e.g.,][]{Frost2018} using standard thermodynamic data tables.  Samples were then quenched to glass by cutting power to the laser. Ambient pressures in the levitation chamber were either 1 bar or $1/3$ bar (see Appendix).  The residual glass spheres were weighed, measured, and in some cases sectioned for inspection prior to dissolution for chemical and isotopic analysis.  Experimental conditions are summarized in Table \ref{Tbl:experiments}.

\subsection{Isotopic and elemental analyses}

Experimental products were crushed into pieces and digested at the Department of of Earth, Planetary and Space Sciences at University of California-Los Angles (UCLA). In preparation for isotopic analyses using a Multi-Collector Inductively Coupled Plasma Mass Spectrometer (MC-ICPMS), Fe and Mg were extracted and purified from the melt residues using ion exchange chromatography in a Class 100 clean wet chemistry laboratory. The samples were digested in Omnitrace HF and HNO$_{3}$ at temperature of ~125 $^o$C for 72-96 hrs on a hot plate. The ion-exchange chromatographic purification procedure for Fe followed \cite{Jordan2019}.  A two-column technique was used for the ion-exchange chromatography purification of Mg after \cite{Young2009}, \cite{Wombacher2009}, and \cite{Tang2021}. Details are provided in Appendix 1.  The isotopic measurements were conducted at UCLA using a ThermoFinnigan Neptune MC-ICPMS instrument. Samples of Fe were run at a mass resolving power ($m/\delta m$) of  $>$8500 to eliminate interferences from $^{40}$Ar$^{14}$N$^{+}$, $^{40}$Ar$^{16}$O$^{+}$, and $\rm{^{40}Ar^{16}O^1H^{+}}$. For Mg isotopic analyses, the isotopic signals were measured  at a mass resolving power of $\sim$6,000. The instrumental fractionation was corrected by using sample-standard bracketing. Uncertainties for each experimental datum are reported as $2\sigma$ internal precision (Appendix 1). Isotope ratio data for USGS standards BHVO-2 and DTS-02 obtained during this study are reported as accuracy checks with $2\sigma$ external precision (Appendix 1). We used IRMM-14 and DSM-3 as the standards for bracketing and as the primary isotopic standards for reporting our Fe and Mg isotope ratios.

The Fe/Al and Mg/Al ratios of the evaporation products  were measured on the MC-ICPMS at UCLA using samples set aside prior to purification at a mass resolving power of  $>$8500.  Elemental and isotope ratios are summarized in Table \ref{Tbl:isotopes}. 
	
\begin{table}
  \centering
  \footnotesize
  \captionsetup{width=0.85\linewidth}
  \caption{Experimental conditions.  D refers to the diameter of the sphere prior to evaporation.  Pressure refers to the pressure in the chamber measured at the capacitance manometer. Ar refers to Ar $>$ CO + CO$_2$ (see Appendix 1). }
    \begin{tabular}{cccccccccc}
    \hline
    \hline
    \head{0.4cm}{Sample} &\head{0.5cm}{D (mm)} & \head{0.6 cm}{initial mass (mg)} &\head{0.6cm}{final mass (mg)} &\head{0.6cm}{$\%$ mass loss}&\head{0.5cm}{time (s)} &\head{0.3cm}{T (K) } &\head{0.4cm}{gas} &\head{1.3cm}{flow rate (cc/min)} &\head{0.6cm}{P (bar)} \\
    \hline
    EL-2.44 & 1.98  & 13.66 & 12.45 & 9.00  & 240   & 2273  & N$_2$    & 360   & $0.33$    \\
    EL-2.45 & 2.11  & 13.63 & 9.42  & 31.50 & 360   & 2273  & N$_2$   & 360   & $0.33$    \\
    EL-2.47 & 2.18  & 14.99 & 12.91 & 13.90 & 270   & 2273  & N$_2$    & 360   & $0.33$   \\
    EL-2.48 & 1.59  & 7.68  & 7.28  & 4.80  & 120   & 2273  & N$_2$    & 284   & $1.0$    \\
    EL-2.60 & 1.72  & 7.28  & 6.03  & 21.60 & 480   & 2273  & N$_2$    & 284   & $1.0$    \\
   {EL-2.73} & 2.15  & 13.86 & 13.05 & 5.80  & 120   & 2273  & N$_2$    & 360   & $0.33$   \\
    EL-2.77 & 1.84  & 9.64  & 9.10  & 5.70  & 360   & 2273  & N$_2$    & 360   & $1.0$   \\
    EL-2.80 & 2.11  & 10.20 & 9.78  & 4.10  & 120   & 2273  & N$_2$    & 360   & $1.0$    \\
    EL-2.82 & 1.89  & 9.37  & 8.83  & 5.80  & 220   & 2273  & N$_2$   & 360   & $1.0$   \\
    EL-2.85 & 1.81  & 9.36  & 8.91  & 4.80  & 240   & 2273  & N$_2$    & 360   & $1.0$    \\
    EL-2.11 & 2.00 & 13.78 &12.97 & 5.87 & 180 & 2323  &Ar & 389 & 1.0 \\
    EL-2.13 & 2.10 & 13.79 & 13.16 & 4.6 & 180 & 2223 & Ar & 137 & 1.0 \\
    EL-2.25 & 2.34 & 19.32 & 18.36 & 4.9 & 200 & 2323 & Ar & 217 & 1.0 \\
    EL-2.26 & 2.22 & 17.30 & 16.29 & 5.8 & 129 & 2273 & Ar & 256 & 1.0 
    \end{tabular}%
  \label{Tbl:experiments}%
\end{table}%

\begin{table}
\begin{adjustwidth}{-0.2in}{-0.2in}
  \footnotesize
  \caption{Isotope results.  Mg isotope ratios are relative to the DSM-3 reference.  EL-mix refers to the starting material for the experiments listed in Table \ref{Tbl:experiments}.  Mg/Mg$_\mathrm{o}$ and Fe/Fe$_\mathrm{o}$
  are obtained from measurements of Mg/Al and Fe/Al, respectively (see Appendix 2).  Data for USGS standards BHVO-2 and DTS-02 are averages and errors in the means, representing external reproducibility.  Uncertainties for experimental products represent internal precision (see Appendix 1).}
    \begin{tabular}{cccccccccccc}
    \hline
    \hline
    Sample &Mg/Mg$_\mathrm{o}$&$\delta ^{25}$Mg &$2 \sigma$ &$\delta ^{26}$Mg& $2 \sigma$ &Fe/Fe$_\mathrm{o}$ & $\delta ^{56}$Fe &$2 \sigma$ &$\delta ^{57}$Fe& $2 \sigma$ \\ 
    \hline
    BHVO-2&-&-0.297&0.007&-0.534&0.010&-&0.128&0.004&0.184&0.006 \\
    DTS-02& - &-0.114&0.006&-0.230&0.010&-&-&-&-& - \\
    EL-mix & 1.000 & -2.791 & 0.010 & -5.453 & 0.021 & 1.000 & 0.385 & 0.025 & 0.564 & 0.043 \\
    EL-mix &  1.000 & -2.790 & 0.011 & -5.443 & 0.021 & 1.000 & 0.385 & 0.011 & 0.564 & 0.019 \\
    EL-2.44 & 1.000 & -2.749 & 0.019 & -5.360 & 0.022 & 0.892 & 0.798 & 0.012 & 1.186 & 0.033 \\
    EL-2.45  &0.951 & -2.420 & 0.018 & -4.726 & 0.030 & 0.455 & 3.992 & 0.011 & 5.921 & 0.019 \\
    EL-2.47  &0.989 & -2.695 & 0.028 & -5.264 & 0.049 & 0.784 & 1.270 & 0.014 & 1.855 & 0.035 \\
    EL-2.48 & 0.999 & -2.755 & 0.022 & -5.364 & 0.020 & 0.937 & 0.649 & 0.021 & 0.974 & 0.032 \\
    EL-2.60 &  0.958 & -2.544 & 0.036 & -4.967 & 0.059 & 0.656 & 2.124 & 0.017 & 3.151 & 0.032 \\
   {EL-2.73} &  1.000 & -2.761 & 0.037 & -5.392 & 0.081 & 0.948 & 0.601 & 0.022 & 0.901 & 0.027 \\
    EL-2.77 &  0.977 & -2.665 & 0.028 & -5.210 & 0.056 & 0.966 & 0.542 & 0.026 & 0.824 & 0.039 \\
    EL-2.80 & 1.000 & -2.758 & 0.032 & -5.387 & 0.064 & 0.992 & 0.417 & 0.012 & 0.644 & 0.052 \\
    EL-2.82 &1.001 & -2.777 & 0.016 & -5.404 & 0.033 & 0.995 & 0.463 & 0.015 & 0.677 & 0.033 \\
    EL-2.85 & 0.991 & -2.740 & 0.015 & -5.328 & 0.035 & 0.958 & 0.477 & 0.010 & 0.715 & 0.036 \\
    EL-2.11 & -  & -2.737 & 0.023 & -5.336 & 0.027& 0.911 & 0.733 & 0.013 & 1.081 & 0.015 \\
    EL-2.13 &  - &   -2.758 & 0.009 & -5.389 & 0.014 & 0.927 & 0.560 & 0.011 & 0.839 &  0.039  \\
    EL-2.25& - & -2.746 & 0.010 & -5.358 & 0.012 & 0.920 & 0.665 & 0.019& 0.964 & 0.044 \\
    EL-2.26& - & -2.765 & 0.016 & -5.385 & 0.020 & 0.902 & 0.672 & 0.016 & 1.005 & 0.039  
    \end{tabular}%
  \label{Tbl:isotopes}%
  \end{adjustwidth}
\end{table}%

\section{Results}

\begin{figure}
\centering
\captionsetup{width=0.95\linewidth}
 \includegraphics[width=5.0in]{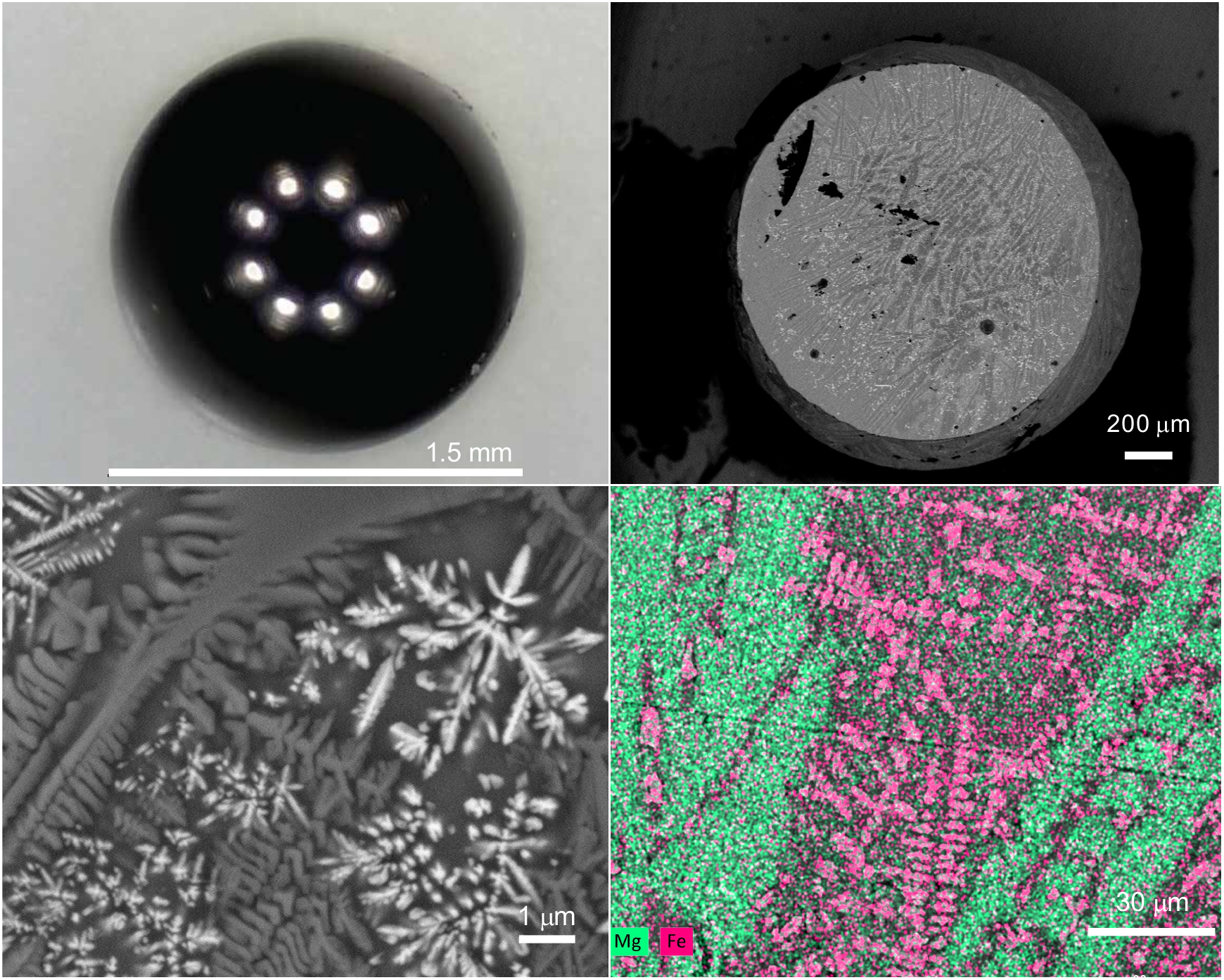}
\caption{Four images of sample EL2.6 post laser heating.  The upper left is an optical image showing the glassy bead prior to dissection. Bright spots are reflections of the ring light on the curved surface of the sphere.  In the upper right is a backscattered electron (BSE) image of a flat cut through the sphere.  The lower left panel shows a closeup BSE image of dendritic quench crystals formed upon rapid cooling.  The lower right panel shows a characteristic x-ray map confirming the presence of quench dendrites of iron oxide, as implied by the BSE images, as well as quench crystals of Mg-rich olivine (green areas) all surrounded by interstitial glass.} 
\label{Fig:semimages} 
\end{figure}

Our experiments are conducted at temperatures more than 600 degrees above the liquid temperature of 1607 K as determined for our initial composition using the Melts program \citep{Ghiorso2002}.  At these temperatures we expect both diffusion and convection to keep the melts well mixed.  As expected, glassy spherules produced by the heating experiments are homogeneous at the mm to tens of $\mu$m scale (see Supplementary Material ).  At the $\sim 1 \mu {\rm m}$  scale dendritic quench crystals of iron oxide are clearly visible (Figure \ref{Fig:semimages}).  The sample shown in Figure \ref{Fig:semimages}, EL2.60, lost $34\%$ of its iron during evaporation (Table \ref{Tbl:experiments}).  We  conclude that the dendrites are quench phases not present during evaporation, and that the molten spheres were well mixed during the evaporation process.  We can therefore describe the evaporation as a Rayleigh process.  A description of the thermodynamic constraints on the evaporation process afforded by these data is given in Appendix 2.  

We use our isotope ratio data to derive gas/melt isotope fractionation factors, $\alpha$, where $\alpha =  (n'/n)_{\rm gas}/(n'/n)_{\rm melt}$ and $(n'/n)$ is the molar isotope ratio for the heavy and light masses of interest, $m'$ and $m$, respectively,  for the indicated phase (e.g., \Feratio\, or \Mgratio\,  in this application).  For our analysis of the results, we express isotope ratios using the delta notation where $\delta = 10^3(({n'/n})/{(n'/n})_{\rm o}-1)$ and $(n'/n)_{\rm o}$ is the initial isotope ratio in the melts prior to evaporation.  Since the melts are well mixed during the experiment, we consider the loss of evaporated species as a Rayleigh fractionation process.  The Rayleigh equation for isotope fractionation in the residual melts as a result of evaporation is

\begin{equation}
\frac{n'/n}{(n'/n)_{o}}=f^{\alpha-1}
\label{Eqn:rayleigh}
\end{equation}
where $f$ is the fraction of the element of interest remaining, or more precisely, the fraction remaining of the major isotope comprising the denominator in $(n'/n)$ (e.g., $^{54}\mathrm{Fe}$ or $^{24}\mathrm{Mg}$ here). Rearranging, we obtain

\begin{equation}
\ln \left(\frac{(n'/n)}{(n'/n)_o}\right) = (1-\alpha)(-\ln f).
\label{Eqn:line}
\end{equation}
Equation \ref{Eqn:line} is that of a line with a slope of $1-\alpha$.  Since $10^3\ln((n'/n)/(n'/n)_{\rm o}) \sim \delta - \delta_o$ is a good approximation over the ranges of isotope ratios reported here, we can plot $\delta-\delta_o$ vs. $-\ln f$ for our experiments and make use of $\alpha = 1-\mathrm{slope}/10^3$ to obtain the effective fractionation factors defined by the data. The method for obtaining values for $f$ for Fe and Mg, or Fe/Fe$_{\rm o}$ and Mg/Mg$_{\rm o}$, respectively, from measured Fe/Al and Mg/Al ratios are described in detail in Appendix 2.  Results are shown in Figure \ref{Fig:isotoperesults}.

\begin{figure}
\centering
\captionsetup{width=0.99\linewidth}
 \includegraphics[width=5.5in]{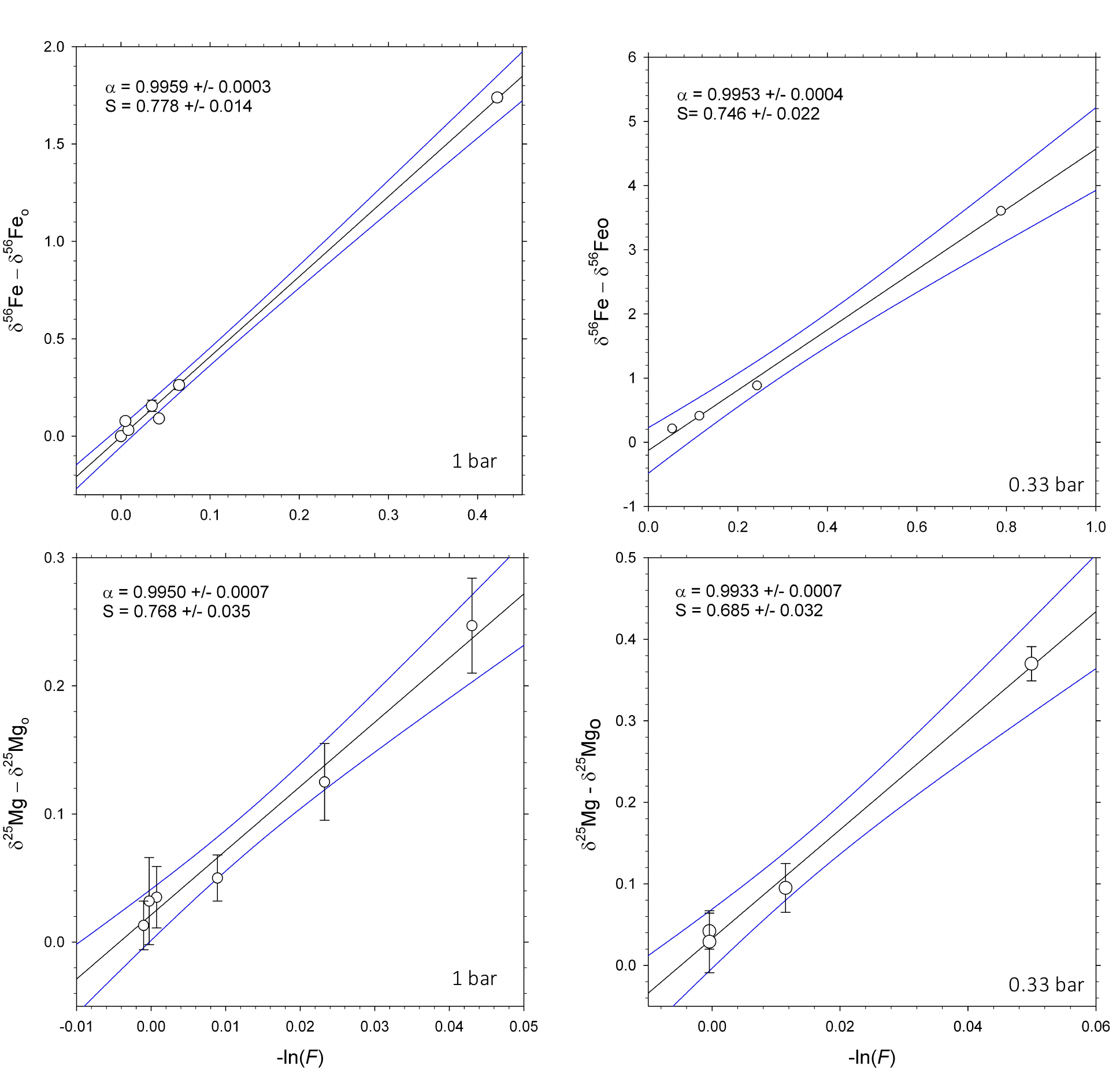}
\caption{Plots of shifts in \Feratio\, and \Mgratio\, vs fraction of iron and magnesium remaining, respectively.  The slopes of the linear best fits to the data yield the isotope fractionation factors and saturation values indicated.  The calculations are described in the text.  All of these data were collected with N$_2$ as the levitation gas.  } 
\label{Fig:isotoperesults} 
\end{figure}

The gas/melt isotope fractionation factors we obtain from the Rayleigh equation reflect the degree of saturation S. A simple relationship between the derived fractionation factor $\alpha$ and S is

\begin{equation}
\Delta = \Delta_{\rm sat} + (1-\mathrm{S})\Delta_{\rm kin},
\label{Eqn:bigdelta}
\end{equation}
where $\Delta = \delta_{\rm gas} - \delta_{\rm melt}$ and is related to the fractionation factor $\alpha$ by the approximation $\Delta \sim 10^3\ln (\alpha)$.  Here the purely kinetic fractionation $\Delta_{\rm kinetic}$ corresponds to the kinetic fractionation factor $\alpha_{\rm kin}$ where $\alpha_{\rm kin} =\sqrt{m/m^\prime}$ and $m^\prime$  is the mass for the heavy isotopic species (Graham's law).  Where S $=0$, the fractionation is purely kinetic.  Where S $=1$, the fractionation corresponds to that at equilibrium. In order to evaluate S using Equation \ref{Eqn:bigdelta} we use the equilibrium isotope fractionation factor for \Feratio\, between solid  FeO and atomic Fe in the vapor from \cite{Polyakov2000-849}.  At the melt temperatures in these experiments, this yields $\Delta_{\rm sat}=-0.088$ \textperthousand .  Similarly, we use the fractionation between Mg$_2$SiO$_4$ and atomic Mg reported by \cite{Schauble2011} for the equilibrium fractionation between the silicate melt and gas phase, yielding $\Delta_{\rm sat} = -0.272$ \textperthousand. 

Our results can be placed in context by comparisons with the first-order expectations for either free evaporation or equilibrium between melt and vapor, comprising two endmember circumstances. The former would be expected if the flow of gas across the surfaces of the evaporating spheres was nearly perfectly effective in removing molecules as they evaporate from the surface.  The latter would be expected if the flow was ineffective in removing evaporation products, permitting a return flux that that in principle could balance the evaporation flux.  The two endmember circumstances correspond to saturation values, S, of $0$ and $1$, respectively.  A first-order estimate of saturation comes from comparing the rate of evaporation to the rate of return flux where the return flux is limited by diffusion through the gas.  In this case \citep{Richter2002-521,Young2019}, 

\begin{equation}
\mathrm{S}=1-\frac{1}{  \left( 1+ \frac{\gamma_i\, r}{D_i} \sqrt{\frac{RT}{2\pi\, m_i}} \right)  },
\label{Eqn:idealS}
\end{equation}
where $r$ is the radius of the evaporating sphere, $D_i$ is the diffusivity of species $i$ in the gas, $m_i$ is the molecular mass, or reduced mass, per mole, $R$ is the gas constant, $T$ is the temperature of the gas, and $\gamma_i$ is the phenomenological evaporation coefficient for $i$.  For the pressures of these experiments, diffusivities are on the order of $10^{-5}$ $\mathrm{m}^2/\mathrm{s}$ \citep{Young2019}. Evaporation coefficients for these elements and these compositions are generally $> 0.05$, and usually on the order of $0.1$ to $0.2$ \citep{Fedkin2006,Schaefer2004}.  Therefore,  values for S predicted from Equation \ref{Eqn:idealS} are  $> 0.992$, much greater than our measured values of $\sim 0.7$ (Figure \ref{Fig:isotoperesults}).    

The gas transport P\'{e}clet number might be used to evaluate the prospects for lowering saturation from the values near unity implied by the pressures of about a bar in the experiments.   The P\'{e}clet number expresses the ratio of advection to gas-phase diffusion, or $uL/D$, where $u$ is the gas velocity near the surface, $L$ is the length scale for the diffusive transport, and $D$ is the diffusion coefficient for molecules in the gas phase.  However, precise values for Pe are difficult to evaluate in this application given uncertainties in the relevant boundary layer thickness, $L$.   For example, at the pressures of these experiments, the mean free path, $\lambda$, is on the order of $10^{-8}$ meters, and if the boundary layer of interest is on the order of several $\lambda$, $L/D$ is $\sim 0.001$, suggesting that gas velocities would have to be on the order of $10^3$ m/s to achieve  Pe$=1$, and velocities of order $10^4$ m/s would be required for Pe on the order of $10$ where advection could be said to dominate over diffusion.  Such  gas velocities are improbably high (see \S \ref{Chapter:Model}).  Given the uncertainties surrounding how to calculate Pe, a different approach is warranted to evaluate the efficacy of gas flow in controlling S.

\section{Model for evaporation of a sphere in flowing gas}{\label{Chapter:Model}

We seek an explanation for the isotope fractionation observed in our experiments.  With apparent saturation values of  $\sim 0.7$ for both Fe and Mg, the results conform  neither to free evaporation where S$ =0$, nor to evaporation at the relatively high pressures of  $\sim 1$ bar where S should be near $1$. This implies that it is the details of the flow regime that govern the extent of saturation. Our solution will be more robust if we restrict our analysis largely to the macroscopic aspects of the experiment rather than on irresolvable boundary layer effects.  However, the nature of the flow affects the interpretations of our results.  In particular, we show below that treating the flowing gas as an inviscid fluid best fits the data.  

\begin{figure}
\centering
\captionsetup{width=0.95\linewidth}
 \includegraphics[width=5.0in]{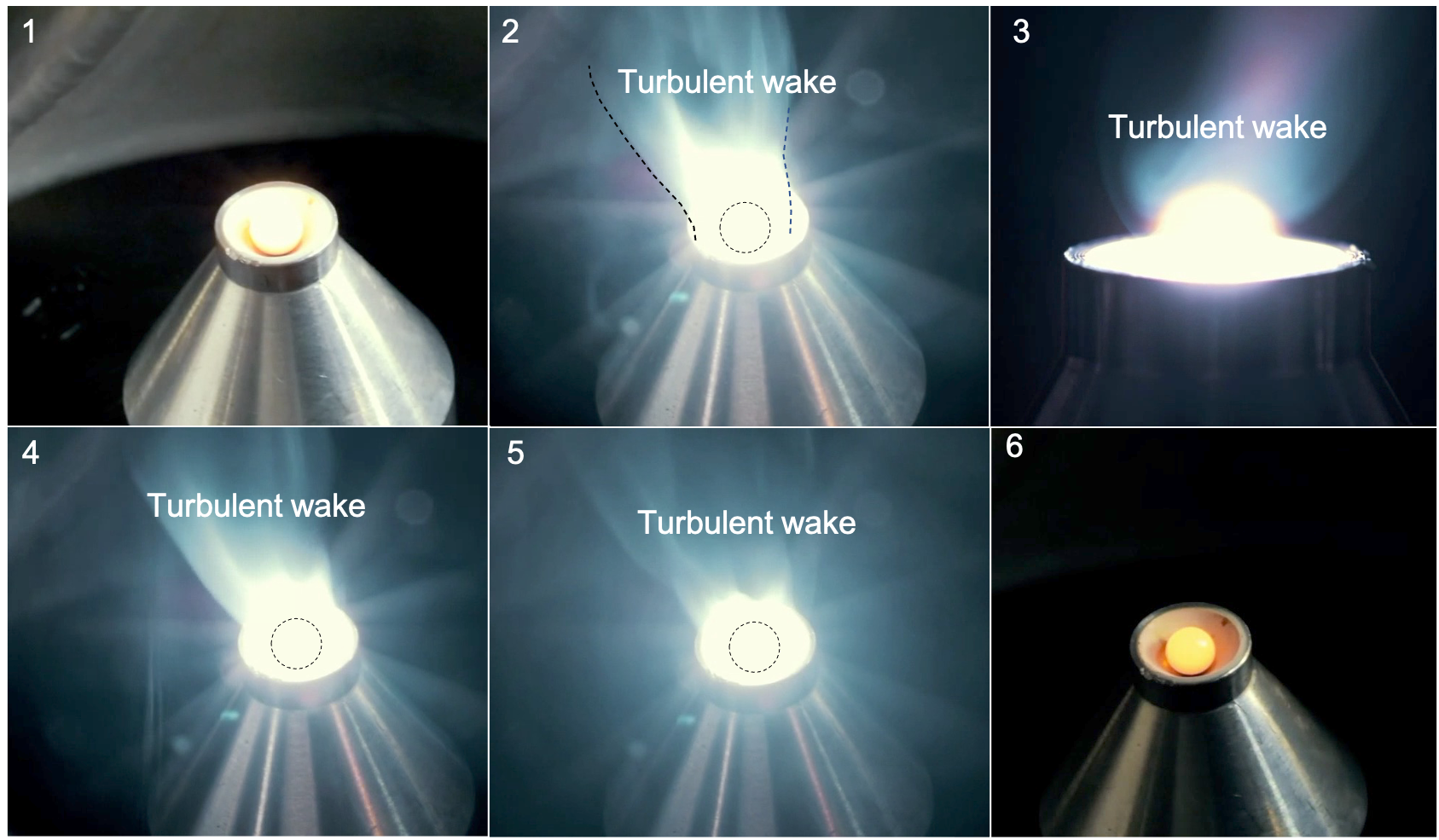}
\caption{Sequence of images taken from a movie of evaporation of a molten sphere in the levitation system under conditions comparable to our experiments.  Flow separation and formation of a turbulent wake at the top of the sphere is evidence for high Reynolds numbers, Re.  The sequence starts with initial heating in the upper left. The wake in image 2 and the position of the sphere in images 2, 4, and 5 are marked by dashed lines.}
\label{Fig:movie}
\end{figure} 

 The flow around the sphere (Figure \ref{Fig:movie}) exhibits flow separation near the equator of the sphere and a prominent turbulent wake.  These features are characteristic of flow at high Reynolds numbers, Re ($=u2r/\nu$ for gas velocity $u$, sphere radius $r$, and kinematic viscosity $\nu$).  At high values for Re of order several thousand or more,  viscosity and surface drag become less important, and the flowing fluid (gas in this case) exhibits more inviscid behavior.  Under these conditions, shear near the surface of the molten sphere that causes frictional drag is subordinate to form drag (pressure effects) \citep[e.g.,][]{Feng2001}.  

We note that the velocity of an inviscid gas flowing across the surface of a sphere and the angle $\theta$ between the surface normals for each point on the sphere and the flow direction is

\begin{equation}
u=V_d(1+1/2) \sin{\theta},
\label{Eqn:thetaflow}
\end{equation}

\noindent where $V_d$ is the free stream velocity of gas unimpeded by the surface of the sphere, referred to here as the drift velocity.  Equation \ref{Eqn:thetaflow} shows that the velocity of the gas, $u$, over the surface of the sphere is greatest  for $\theta = \pi/2$, a manifestation of the Bernoulli effect, and in principle would vanish at the stagnant points at the poles of the sphere coinciding with the flow direction \citep{McDonald2015} in the case of an ideal inviscid fluid  (Figure \ref{Fig:schematic}).  This increase in velocity on the sides of the sphere roughly parallel to the overall flow direction stabilizes the levitated spheres. 

\begin{figure}
\centering
\captionsetup{width=0.95\linewidth}
 \includegraphics[width=5.0in]{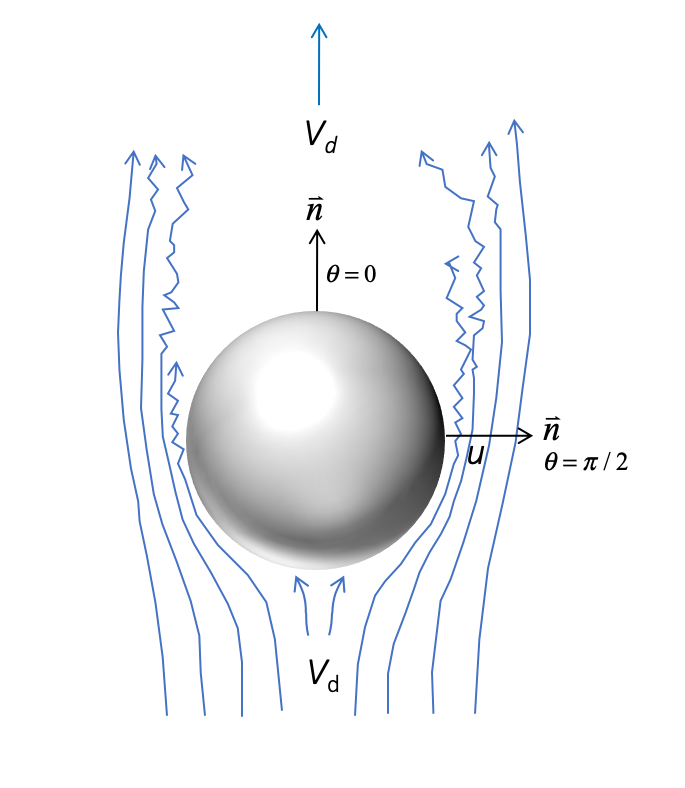}
\caption{Schematic diagram showing the flow field around the evaporating sphere.  Flow separation at the top of the sphere is evidence for high Reynolds numbers.  Velocity $V_d$ is the drift velocity, or free stream velocity, described in the text.  Velocity $u$ is that directly adjacent the sphere.  Angle $\theta$ is between the gas flow and the surface normals, $\vec{n}$. }
\label{Fig:schematic}
\end{figure} 

We derive the free stream velocity, or drift velocity, from the force balance that leads to stability during aerodynamic levitation. The opposing forces leading to stable levitation are gravity acting on the mass of the sphere, $gm_{\rm sphere}$, and the drag force $f_{\rm drag} = 4\pi r^2 (C_D/2) \overline{\rho}_{\rm gas}  V_d^2$.  Solving this force balance equation for gas velocity  leads to    

\begin{equation}
V_d=\frac{  \sqrt{2 g\, m_{\rm sphere}  } } { 2\sqrt{\pi C_D \overline{\rho}_{\rm gas}}   } \frac{1}{r}.
\label{Eqn:drag}
\end{equation}\

\noindent Because $C_D$ depends on Re,  Re  depends on $V_d$, and $V_d$ depends on $C_D$, self consistent solutions are obtained for viscosity, Reynolds number, and velocity by iteration.   The relationship between $C_D$ and Re for flow of an inviscid fluid characterized by a no-shear boundary condition, as opposed to a no-slip boundary condition, was investigated by \cite{Moore1963}. The drag coefficient under these circumstances as determined by Moore is

\begin{equation}
C_D=\frac{48}{\mathrm{Re}}
\left(1-\frac{2.2}{\mathrm{Re}^{1/2}}\right)
\label{Eqn:Drag_Moore}
\end{equation}\

\noindent where here  $\mathrm{Re}=2V_d r/\nu$ for gas drift velocity $V_d$.  We calculated the gas drift velocities corresponding to a finite kinematic viscosity, $\nu$, for N$_2$  gas based on the model of  \cite{Lemmon2004} and the relationship between the drag coefficient, $C_D$, and the Reynolds number from Equation \ref{Eqn:Drag_Moore}.  The solution  yields $V_d = 41.7$ m/s, Re $= 5180$, and $\nu = 1.61\times 10^{-5}$ m$^2$/s for an ambient gas pressure of 1 bar, an ambient temperature of 298K, a gas density of 1.13 kg/m$^3$, $g = 9.8$ m/s$^2$, a molten sphere density of 2700 kg/m$^3$, and a sphere radius of $1.0\times 10^{-3}$ m .  For an ambient gas pressure of $0.333$ bar, we obtain $V_d = 43.3$ m/s under the same conditions.  

This gas velocity is greater than that previously estimated by \cite{Badro2021} for broadly similar experimental conditions by a factor of $\sim 4$.  The reason is that we use the drag coefficient suitable for a no-shear boundary rather than that for a no-slip boundary.  As shown below, the higher velocities implied by this inviscid behavior of the gas are consistent with the observed isotope fractionations while the assumption of a no-slip boundary condition offers no clear explanation for the observed  fractionations.  

 This gas velocity $V_d$ has the effect of either working against, or aiding, the return of evaporated molecules to the melt surface, depending upon the angle $\theta$ of the normal to the surface of the sphere relative to the overall flow direction of the gas.  This return flux mitigates the loss of molecules from the melt by evaporation.  Next we calculate this effect and apply it to the experiments.

The Hertz-Knudsen equation for free evaporation provides the means to calculate the expected evaporation flux that is balanced to more or lesser degrees by the return flux to the surface of the melt.  It is useful for what follows to show that the equation  is obtained  from the Maxwell-Boltzmann distribution of velocities where the frequency distribution for velocities centered at a mean velocity of 0 is given by

\begin{equation}
f(V_z,0) =\frac{   1 }{   V_t\sqrt{\pi}} \exp\left({\frac{  -(V_z-0)^2}{  V_t^2}}\right),
\label{Eqn:maxwell}
\end{equation}
and where $V_z$ represents velocities along the $z$ direction and $V_t$ is the thermal velocity representing the dispersion of velocities about the mean  given by 

\begin{equation}
V_t=\left(\frac{2RT}{m}\right)^{1/2},
\label{Eqn:thermalvelocity}
\end{equation}
where $m$ is the molecular mass per mole.  We adopt the surface of the melt as the location of the center of the velocity distribution perpendicular to the melt surface. The evaporative flux of molecules is obtained from the product of the gas number density at saturation, $\rho_{\rm sat}$, and  gas velocity using the integral of the Maxwellian velocity distribution in Equation \ref{Eqn:maxwell} from $0$ to $+\infty$, yielding

\begin{align}
F_{\rm evap}=& \rho_{\rm sat} \int_{0}^{\infty} \frac{  V_z }{   V_t\sqrt{\pi}} \exp\left({\frac{  -(V_z-0)^2}{  V_t^2}}\right)\, dV_z \nonumber  \\[10pt]
=& \frac{   P_{\rm sat}  }{ RT } \frac{ V_t}{2 \sqrt{\pi}},
\label{Eqn:evapflux}
\end{align}\

\noindent
where  we have substituted $P_{\rm sat}/(RT)$ for the gas number density at saturation.  Substitution of Equation \ref{Eqn:thermalvelocity} into \ref{Eqn:evapflux} results in the Hertz-Knudsen equation for free evaporation as commonly written:

\begin{equation}
F_{\rm evap}=\frac{   P_{\rm sat}}{  \sqrt{2\pi m RT}}.
\label{Eqn:hertzknudsen}
\end{equation}\

\noindent It is convenient to consider the surface-integrated evaporated flux, written as

\begin{align}
F_{S,{\rm evap}}=&\int_{S} \frac{   P_{\rm sat}}{  \sqrt{2\pi m RT}}\, dS \nonumber \\[10pt]
=& 4\pi r^2 \frac{   P_{\rm sat}}{  \sqrt{2\pi m RT}}
\label{Eqn:surfaceevap}
\end{align}\

\noindent
where in this case $S$ is the surface of the molten sphere and we consider that evaporation occurs uniformly over the entire surface.  
A similar calculation but integrating the velocity distribution from $-\infty$ to $0$ results in the return flux to the surface of the evaporating melt written in terms of the ambient pressure $P$ rather than the saturation pressure:

\begin{equation}
F_{S,{\rm return}}=4\pi r^2 \frac{  - P}{  \sqrt{2\pi m RT}}.
\label{Eqn:surfacereturn}
\end{equation}\

Charnoz et al. (\citeyear{Charnoz2021}) showed that the effect of a unidirectional flow of gas on the return flux  can be accommodated by modifying the Maxwell-Boltzmann velocity distribution so that the velocity distribution is centered on the average drift velocity of the moving gas.  We adopt this same general approach here, but we modify it to account for the spherical geometry of the evaporation surface that results in asymmetry in the flow effects as a function of the polar angle $\theta$ to the flow direction (but symmetric in the azimuthal angle).  Following the same general approach as that used to derive the surface-integrated evaporation and return fluxes in the absence of flow, the return flux normal to the evaporation surface along the flow direction $z$ is

\begin{align}
F_{\rm return}=& \rho_{\rm gas} \int_{-\infty}^{0} \frac{  V_z }{   V_t\sqrt{\pi}} \exp\left({\frac{  -(V_z-V_d)^2}{  V_t^2}}\right)\, dV_z \nonumber  \\[10pt]
=& \rho_{\rm gas} \left( \frac{   V_d \, \, \mathrm{erfc}(V_d/V_t)}{ 2} - \frac{ V_t \, \exp(-V_d^2/V_t^2)}{ 2\sqrt{\pi}} \right),
\label{Eqn:returnwithflow}
\end{align}\

\noindent
where $\rho_{\rm gas}$ is the density of the gas above the surface equal to $P/(RT)$.  The surface-integrated return flux requires accounting for the changing angle between the flow direction and the surface normals to the sphere $\vec{n}$.  The return flux at the top of the levitated sphere in our application is opposite to the direction of the flowing gas.  Conversely, on the sides of the sphere, the gas flow is perpendicular to the normal to the evaporation surface, and the return flux approaches zero.  Upstream, on the bottom of the levitated sphere, the return flux is in the direction of flow.  This variation necessitates expressing $F_{\rm return}$ as a vector projected onto the normals to the surface of the sphere, yielding

\begin{equation}
\vec{F}_{\rm return}\cdot \vec{n}=\rho_{\rm gas} \left( \frac{   V_d \cos{\theta} \, \, \mathrm{erfc}(V_d \cos{\theta}/V_t)}{ 2} - \frac{ V_t \exp(-(V_d \cos{\theta})^2/V_t^2)}{ 2\sqrt{\pi}} \right),
\label{Eqn:dotproduct}
\end{equation} \

\noindent
where again, $\theta$ is the angle between the gas flow direction and the normal to the surface of the sphere.  Integrating over the surface of the evaporating sphere, using

\begin{equation}
F_{S,{\rm return}}=4 r^2 \int_{\phi=0}^{\pi/2} \int_{\theta=0}^{\pi} \vec{F}_{\rm return}\cdot \vec{n}\,  \sin{\theta} \, d\theta \, d\phi
\label{Eqn:surfacintegral}
\end{equation}\

\noindent
results in the surface-integrated return flux to the sphere in the presence of a unidirectional flow of gas with velocity $V_d$

\begin{equation}
F_{S,{\rm return}} = \rho_{\rm gas} \, 2\pi r^2 \left(  \frac{  -(V_t^2+2V_d^2)\, \mathrm{erf}(V_d/V_t) } {4V_d}- \frac{  V_t \exp{-V_d^2/V_t^2}  }{2\sqrt{\pi}}  \right).
\label{Eqn:surfacereturnwithflow}
\end{equation}

We solve for the saturation of evaporating species in the gas phase, S$=\rho_{\rm gas}/\rho_{\rm sat} = P/P_{\rm sat}$, by noting that the difference between the surface integrated evaporation and return fluxes is the surface-integrated escape flux.  The escape flux is the product of gas density and drift velocity, $F_{\rm escape} = \rho_{\rm gas} V_d$.  Mass balance therefore requires that

\begin{align}
4\pi r^2 \, \rho_{\rm gas} V_d =& 4\pi r^2\, \rho_{\rm sat} \frac{ V_t}{2 \sqrt{\pi}} +  \nonumber \\[10pt]
& \rho_{\rm gas} \, 2\pi r^2 \left( \frac{   -(V_t^2+2V_d^2)\, \mathrm{erf}(V_d/V_t)  } { 4V_d} - \frac{  V_t \exp{-V_d^2/V_t^2}  }{ 2\sqrt{\pi}  } \right),
\label{Eqn:fluxbalance}
\end{align}
which can be solved for $\rho_{\rm gas}/\rho_{\rm sat} $, yielding

\begin{equation}
{\rm S}=
\frac{2\sqrt{\pi} V_t }
{
4\pi V_d  + 2\pi \left( \frac{ (V_t^2+ 2V_d^2)\, \mathrm{erf}(V_d/V_t)}{4V_d} + \frac{V_t \exp(-V_d^2/V_t^2)} {2\sqrt{\pi}}\right)
}.
\label{Eqn:saturation}
\end{equation}\

\noindent Equation \ref{Eqn:saturation} is plotted in Figure  \ref{Fig:saturation} in which saturation is plotted against the ratio of flow velocity to thermal velocity.  Because the limit of the right hand side of Equation \ref{Eqn:saturation} is 1 as $V_d$ goes to $0$, it scales the saturation relative to the case where $V_d = 0$. In other words, if the r.h.s.\  of  Equation \ref{Eqn:saturation} is defined as $\Omega$,  in general we have $\mathrm{S}(V_d) = \mathrm{S}(0) \Omega$. 
For an ideal, inviscid gas, $\Omega$ is independent of pressure (Figure \ref{Fig:saturation}).

\begin{figure}
\centering
\captionsetup{width=0.90\linewidth}
 \includegraphics[width=3.3in]{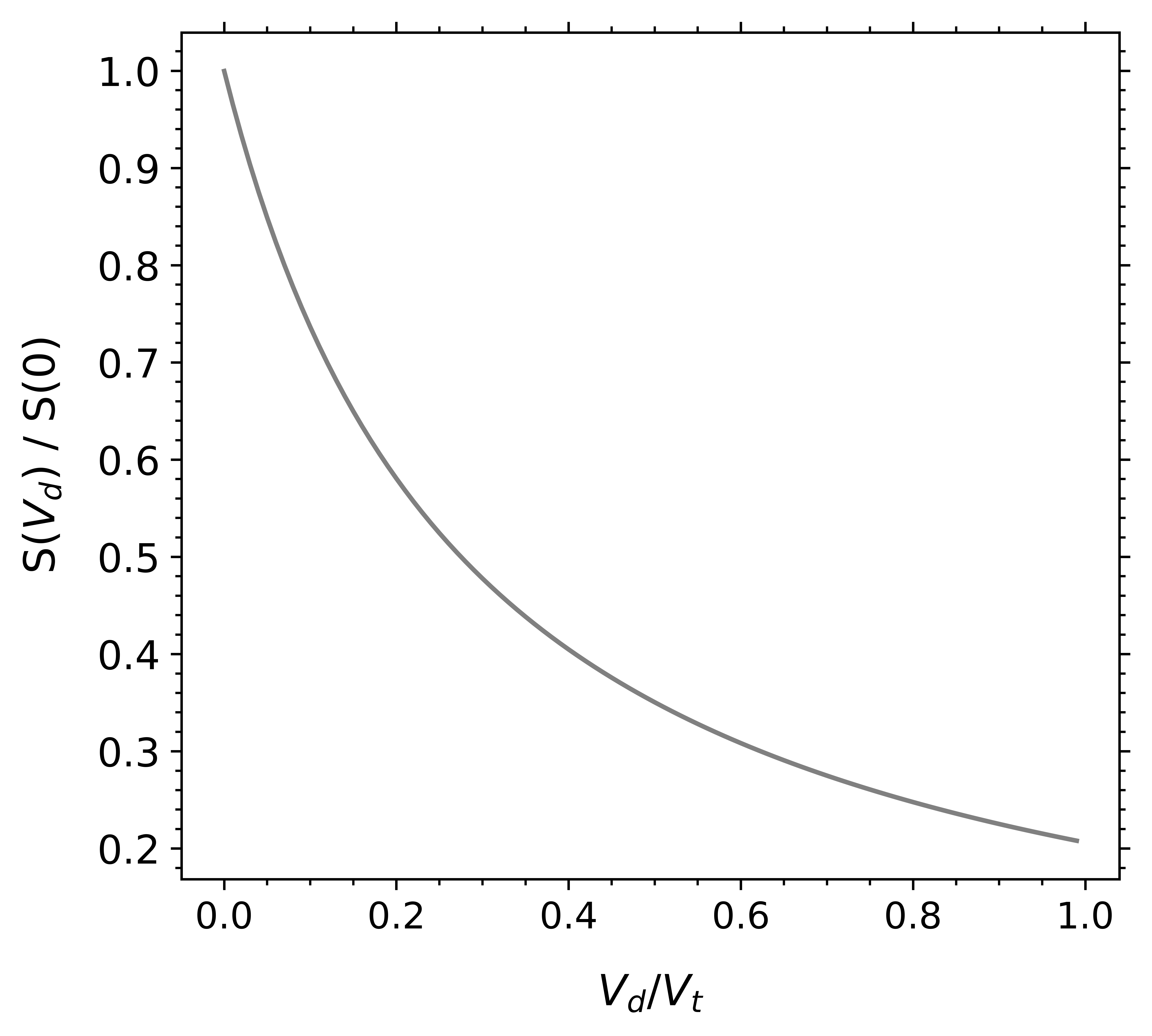}
\caption{Plot of saturation S($V_d$)/S($0$) as a function of the ratio of gas flow velocity to thermal velocity based on Equation \ref{Eqn:saturation}.  The plot can be used to estimate the departure from a given saturation due solely to the drift velocity $V_d$ relative to the thermal velocity $V_t$ for gas flowing past an evaporating sphere. } 
\label{Fig:saturation} 
\end{figure}

\section{Discussion}

\subsection{Application of the model  to experiments}

We apply the model described in \S \ref{Chapter:Model} to our experiments to compare the predicted saturation to the saturation values derived from the observed isotope fractionation. The critical input is the $V_d/V_t$.  For evaporation of Fe into an N$_2$ gas with a temperature of $298$ K at a gas pressure of $1.0$, the corresponding levitation gas velocity $V_d$ of $41.7$ m/s and thermal velocity $V_t$ of $515.4$ m/s, yields a $V_d/V_t$ ratio of $0.081$. The value of $0.081$ for $V_d/V_t$ corresponds to a saturation of $0.776$ (Figure \ref{Fig:saturation}, Table \ref{Tbl:model}).  Repeating the calculation for a pressure of 0.333 bar yields a $V_d/V_t$ ratio of $0.084$ and a saturation of $0.769$. We use the ambient temperature of the gas in this calculation because of the high stream velocity relative to the thermal diffusivity of the gas (Appendix 3).    

Our predicted range in saturation from $0.776$ to $0.769$ using N$_2$ as the levitation gas  compares favorably with our experimental \Feratio\, saturation values of $0.778\pm 0.014$ and $0.746 \pm 0.022$ for chamber pressures of 1 bar and $0.333$ bar, respectively.  The model calculations are relatively insensitive to the precise pressure in the range from 2.4 bar to 0.333 bar (the full conceivable range of ambient pressures).  There is a prediction for differences in the calculated saturation values between evaporation of Fe and Mg due to our use of the reduced masses in the calculation of thermal velocities. For Mg isotopes, we calculate saturation values of $0.806$ and $0.800$ under these same conditions, respectively.  These also compare reasonably well with the measured saturation values of $0.768\pm 0.035$ and $0.685\pm 0.032$ for \Mgratio\, at chamber pressures of 1 bar and $0.333$ bar, respectively.  

We point out that the precise saturation values we derive from our experiments are based on the assumption of the applicability of Graham's law.  In cases where the kinetic fractionation factor is closer to unity  than predicted by Graham's law \citep[e.g.,][]{Richter2007}, our experimental saturation values will be overestimates (Equation \ref{Eqn:bigdelta}).  However, the approach taken here is to construct our analysis on  the basis of the least number of assumptions about the kinetics of the purely evaporative process at the surfaces of the melts, so we confine our analysis to  first-order estimates for kinetic fractionation factors.  

Because of the dependence of thermal velocity and gas density on molecular mass, the model predicts slight differences in saturation with molecular weight of the levitation gas.  Replacing N$_2$ with Ar in our 1.0-bar calculations changes the predicted Fe isotope saturation from $0.776$ to $0.759$.  We have few data for evaporation in flowing Ar.  In the experiments in hand, S $= 0.83 \pm 0.07$, compared with $0.776 \pm 0.014$ for N$_2$ under similar conditions.  The uncertainties associated with the Ar experimental results  makes the differences between the N$_2$ and Ar experiments inconclusive.    

The generally good agreement between the model predictions and the experiments suggests that the levitation gas behaves as an inviscid fluid near the surfaces of the actively evaporating spheres, and that flow of the gas controls the degree of saturation. The geometric factors are important aspects of the model that produce agreement with the experimental results.  For example, if we replace our spherical model with a 1D planar model in which evaporation, return flux, and drift velocity $V_d$ are all along a single axis perpendicular to the planar surface, the predicted saturation values are near 0.870, all else equal, considerably greater than the observed values.

As an alternative to the model based on the inviscid behavior of the gas, we investigated the gas velocities obtained from the kinematic viscosity, $\nu$, of N$_2$  gas from  \cite{Lemmon2004} and the relationship between the drag coefficient, $C_D$, and the Reynolds number appropriate for a no-slip condition at the surface of the evaporating sphere at high Reynolds numbers from \cite{Feng2001}.  The latter assumes that viscous forces dominate near the evaporating surface.  The result for stable levitation of the melt spheres at 1 bar pressure of N$_2$ and a gas temperature of 298 K yields $C_D = 0.554$, Re $ = 496$, and $V_d = 3.995$ m/s.  The slower velocity relative to the thermal velocity corresponds to $\mathrm{S}=0.973$ for Fe isotopes, all else equal.  This higher value for saturation is far removed from the observed values, and further suggests that the levitation gas should be treated as inviscid for our purposes.  We conclude from this analysis that treating the fluid as having minimal drag on the surfaces of the levitated, molten spheres is consistent with the observed isotope fractionation.   

The source of the inviscid behavior is unclear, but may be due to evaporation at the surface.  Evaporation is known to reduce frictional drag on the surfaces of evaporating bodies and possibly form drag as well \citep{Eisenklam1967,Depredurand2010,Kurose2003}.  \cite{Eisenklam1967} showed that experiments illustrating the effect of evaporation on drag can be explained by the relation

\begin{equation}
\frac{C_D}{C_D^*}=\frac{1}{1+B_m},
\label{Eqn:CD_evap}
\end{equation}\

\noindent where $C_D^*$ is the drag coefficient in the absence of evaporation and $B_m$ is the Spalding mass transfer number given by

\begin{equation}
B_m=\frac{Y_s - Y_{\rm amb}}{1-Y_s}.
\label{Eqn:spalding}
\end{equation}

\noindent Here $Y_s$ is the mass fraction of the evaporating vapor in the vapor phase at the surface of the molten sphere and $Y_{\rm amb}$ is the mass fraction of evaporating vapor in the ambient gas.  Values for $Y_s$ can be obtained in terms of the mole fraction of evaporating species for the vapor at the surface, $x_s$,  using

\begin{equation}
Y_s=\frac{x_s MW_i}{x_s MW_i + (1-x_s) MW_{\rm gas}},
\label{Eqn:Ys}
\end{equation}\

\noindent where $MW_i$ is the molecular weight of the gas emitted by evaporation and $MW_{\rm gas}$ is the molecular weight of the ambient gas \citep[e.g.,][]{Depredurand2010}. We can use Equations \ref{Eqn:CD_evap} through \ref{Eqn:Ys} to assess the mass loading at the surfaces of the evaporating spheres due to evaporation that would be required to explain the low drag evidenced by the isotope results.  In these experiments, $Y_{\rm amb} = 0$, $MW_i \sim 48$ (e.g., average of Fe + 1/2O$_2$), $MW_{\rm gas} = 28$, and $C_D/C_D^* \sim 0.009/0.5$ based on our calculated drag coefficient for the no-shear boundary condition vs.\ that for a no-slip boundary.  These values require that $x_s = 0.969$, indicating that the inviscid behavior of the gas would be explained if the vapor at the surfaces of the evaporating spheres were composed of more than 95\% rock vapor by mole.  This seems entirely plausible. 

As noted previously by \cite{Badro2021}, the value for S of $\sim 0.75$ has been found in most previous laser-heating aerodynamic levitation experiments in which isotope fractionation was measured, suggesting that it is the aerodynamics of the gas flow over the spheres that controls saturation.  Our model explains all of these previous data for Mg, Si, Fe, and Cu isotope fractionation \citep{Ni2021, Badro2021}, some of which (Cu) were collected using the same apparatus as that used in this study.  The data for Zn isotopes \citep{Wimpenny2019} are complicated by partitioning of Zn into separate phases during the evaporation process.  The resulting episodic losses of Zn yield higher apparent saturation values $> 0.8$.  

Since the low-viscosity behavior observed in these experiments would be expected to be exacerbated for an H$_2$-rich gas because of its low specific gas density, increasing the likelihood for high $V_d/V_t$ values, and suggesting that the model described here should be applicable to various astrophysical environments.  

\begin{table}
\caption{Model saturation values for Fe and Mg isotope saturation.}
\begin{center}
\begin{tabular}{lcccc}
\hline 
\hline
\multicolumn{1}{c}{}& \multicolumn{3}{c}{}\\
Model&S(\Feratio\,)&$V_d/V_t$&S(\Mgratio\,)&$V_d/V_t$\\
\hline
N$_2$, 1.0 bar&$0.776$&$0.081$&$0.806$&$0.068$\\
N$_2$, 0.33 bar&$0.769$&$0.084$&$0.800$&$0.070$\\
Ar, 1.0 bar&$0.759$&$0.089$&$0.797$&$0.072$\\
\end{tabular}
\end{center}
\label{Tbl:model}
\end{table}

\subsection{Application to pebble accretion}

The model that explains the isotope fractionation observed in the laser-heating aerodynamic levitation experiments may be applied to a number of different astrophysical environments where rock material is expected to melt in the presence of enveloping gas.  One such environment is pebble accretion. Pebble accretion is  potentially the primary mechanism for growing rocky planetary cores on timescales comparable to the lifetimes of protoplanetary disks \citep{Ormel2017}.  Once reaching the mass of Mars, the rapidly growing planets can accrue H$_2$-rich atmospheres with masses on the order of a per cent of that of the planet \citep{Lee2016a, Ginzburg2016a, Fressin2013a}.  The sizes of the accreting ``pebbles" depend on the local disk gas densities, but can range from $\mu{\rm m}$ to meters.  Here we consider the fate of pebbles entering the hydrogen-rich atmosphere of a growing planet of mass  $M_{\rm p}=0.5 M_{\oplus}$ and radius of $R_{\rm p}=0.841 R_{\oplus}$, representing  a proto-Earth prior to shedding its primary atmosphere; in general planets with $M_{\rm p}<1.8 M_{\oplus}$ are expected to lose their primary atmospheres \citep[e.g.,][]{Gupta2019a}.    

For illustration purposes, we make use of the simplest possible representation of an atmosphere, composed of a lower, adiabatic (well-mixed) layer overlain by a thinner, radiative layer that we approximate as being isothermal at some prescribed equilibrium temperature.  The temperatures  as a function of elevation $z$ is in this case

 \begin{equation}
 \begin{cases}
T(z)=T_0\left( 1-\frac{\gamma -1}{\gamma} \frac{z}{H(z)}\right) & z < z_{\rm homopause} \\
T(z)=T_{\rm Eq} & z > z_{\rm homopause},
\end{cases}
\label{Eqn:Tatm}
\end{equation}

\noindent where  $z_{\rm homopause}$ is the altitude at the top of the convective layer where the adiabatic temperature equals the equilibrium temperature, $T_{\rm Eq}$, $\gamma$ is the adiabatic index, $7/5$ for our diatomic H$_2$ atmosphere, $T_0$ is the temperature at the base of the atmosphere,   $H(z)=RT(z)/(m_{\rm gas} g(z))$ is the scale height at $z$, $m_{\rm gas}$ is the molar molecular weight of the gas, and $g(z) = GM_{\rm p}/(z+R_{\rm p})^2$.   We use $T_0= 2500$ K, consistent with estimates for surface temperature under these conditions \citep{Ginzburg2016a}.  The surface pressure is estimated from the weight of the overlying atmosphere: $P_ 0= g_0(x_{\rm atm}M_{\rm p})/(4\pi R_{\rm p}^2)$, where $x_{\rm atm}$ is the mass fraction of the planet composed of atmosphere.
 The pressure at each altitude is obtained numerically using

\begin{equation}
 \begin{cases}
d\ln(P(z))=\frac{\gamma}{\gamma -1}d\ln(T(z)) &z < z_{\rm homopause} \\
P(z)=P(z_{\rm homopause})\exp\left(-\frac{(z-z_{\rm homopause})}{H(z_{\rm homopause})}\right)& z > z_{\rm homopause}.
\end{cases}
\label{Eqn:Patm}
\end{equation}

\noindent  Upon entering the atmosphere, pebbles quickly achieve terminal velocity and experience frictional heating.  An estimate of pebble temperature is obtained from the balance between frictional heating and radiation at terminal velocity.  The rate of energy gained by frictional heating, $\dot{E}$, is obtained from the drag force multiplied by velocity, so that $\dot{E}= f_{\rm drag} \times V_d$.  The radiative energy is $\dot{E}= 4\pi r^2 \sigma_{\rm SB}(T^4-T^4_{\rm bg})$ where we assume unit emissivity (results are not sensitive to this assumption), $\sigma_{\rm SB}$ is the Stefan-Boltzmann constant, and $T_{\rm bg}$ is the background temperature of the surrounding atmosphere.  Equating the two expressions for power, we have for the temperature of the pebble at terminal velocity

\begin{equation}
T=\left( T_{\rm bg}^4+\frac{(C_D/2) \bar{\rho}_{\rm gas}}{\sigma_{\rm SB}}V^3_d\right)^{1/4}.
\label{Eqn:grainT}
\end{equation}
Where the atmosphere is hot compared with the effects of frictional heating, $T_{\rm bg}$ determines the temperature of the pebbles. 

 In order to estimate saturation values as a function of elevation in the atmosphere, we require self consistent values for the terminal velocities, Reynolds numbers and drag coefficients.  This requires  kinematic viscosities of the H$_2$ gas calculated from  dynamic viscosities and gas densities at each elevation.  The dynamic viscosities of H$_2$ in this calculation are obtained from the temperature-dependent Sutherland constants for hydrogen gas from \cite{Braun2018} and ideal gas densities corresponding to the atmospheric H$_2$ pressures.  The terminal velocities depend on the diameter of the pebbles.  Our solutions therefore must be coupled to a model for the progressive evaporation of the pebbles as they heat up.   Experimental data for evaporation of forsterite in H$_2$ gas suggest a temperature-dependent evaporation flux far from saturation of $J({\rm moles}/({\rm m^2 s}))=4.8\times 10^{5}\exp({-36082/T})$  where the activation energy (300 kJ/mole) applies to evaporation in the presence of hydrogen \citep{Nagahara1996-1445, Young2000-321,Richter2002-521}.  The rate of mass loss of the pebbles is $\dot{m}=-J A (1-{\rm S}) MW$, where A is the surface area of the evaporating sphere and $MW$ is its molecular weight ($\sim 0.145$ kg/mole).  The rate of evaporation increases dramatically above 2000 K so most of the mass loss will occur above this temperature.  Using this formulation for the rate of evaporation, we calculated numerically the saturation attending partial evaporation of 1 cm pebbles descending through our model atmosphere using the same method used to obtain self-consistent solutions for the experiments described above.   We have not included the cooling effects of the latent heats of melting and evaporation, but note that the latent heats are small compared with frictional heating.   For example, for a typical enthalpy of melting of $\sim 4\times 10^5$ J kg$^{-1}$, the conversion of a 5 mm pebble, a mass of $0.0015$ kg, from solid to melt consumes 600 J.  The frictional heating of this object passing through the atmosphere below the point of melting generates $\sim 7000$ J. Enthalpies of vaporization are a few times larger than those of fusion \citep[e.g.,][]{Costa2017}, but vaporization is likely to be incomplete (see below), suggesting this effect has only a modest effect on the thermal budget.

\begin{figure}
\centering
\captionsetup{width=0.99\linewidth}
 \includegraphics[width=5.7in]{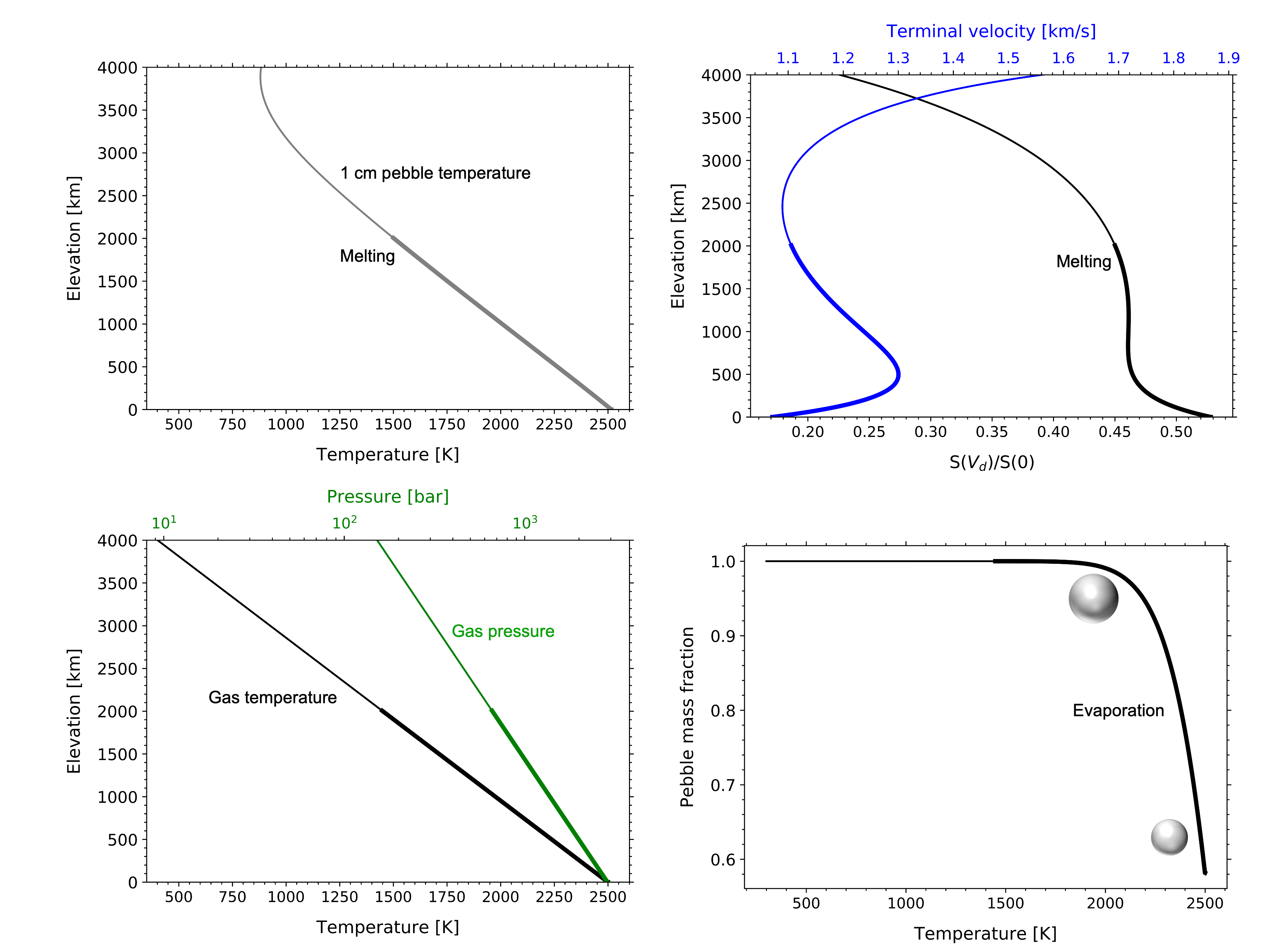}
\caption{Four plots showing the results of ``pebbles" entering a primary atmosphere of H$_2$ surrounding a planet with mass of  $M_p=0.5 M_{\oplus}$ and a radius of $R_p=0.841 R_{\oplus}$, representing  a proto-Earth prior to loss of its primary atmosphere.  The mass of the atmosphere is taken to be $0.5\%$ of the mass of the planet.  Upper left: pebble temperature vs.\ altitude $z$.  Upper right: saturation relative to saturation with zero headwind velocity and terminal velocity vs.\ altitude $z$.  Lower left: atmospheric temperature  and pressure vs. altitude.  Lower right: mass fraction remaining due to evaporation vs.\ temperature, where the spheres depict the shrinking of the pebbles as they descend.  Results are shown for a pebble size of 1 cm. Thicker portions of the curves denote where the pebble is at least partially molten (i.e., where temperatures are above the solidus).
  } 
\label{Fig:pebble_accretion} 
\end{figure}

Results  for 1 cm pebbles entering an atmosphere of H$_2$ comprising $0.5\%$ of the mass of the planet are shown in Figure \ref{Fig:pebble_accretion}.  The equilibrium temperature of the atmosphere is assumed to be $300$ K.  At an altitude just below 2000 km, the grains encounter temperatures of  $\sim 1500$ K, indicating the onset of melting based on the solidus from melting experiments for chondrite meteorites \citep{McCoy1999}.  The temperatures at these depths in the atmosphere arise primarily from coupling between the pebbles and the surrounding hot gas with a lesser component of  frictional heating.  The terminal velocities upon melting are on the order of $\sim 1.1$ to $1.2$ km/s (Figure \ref{Fig:pebble_accretion}).  At these velocities the pebbles require about 30 minutes to complete their fall to the surface after melting begins. The flight time is sufficiently long that pebbles of this size would be completely melted.  The 1 cm pebbles lose approximately $40\%$ of their mass due to evaporation as they descend to the surface (Figure \ref{Fig:pebble_accretion}). 

In this scenario, melting occurs in the ambient atmosphere at temperatures of $> 1500$ K and pressures of $\sim 1000$ bar.  At these pressures, in the absence of a headwind, a sphere of molten rock would be saturated due to the return flux imposed by the total pressure, and no appreciable fractionation of isotopes or elemental abundances would be expected.  However, the plot of S$(V_d)$/S$(0)$ versus altitude shows that for this pebble size, the effect of the velocity of gas over the surface of the melting grain is to lower the saturation from near unity to $\sim 0.45$.  As we have seen from the laser-heating aerodynamic levitation experiments, a saturation this low is more than sufficient for significant isotopic and elemental fractionation. Smaller pebbles than those shown here have lower terminal velocities and are closer to saturation (e.g., S for a 5 mm pebble is approximately 0.80) while larger pebbles have higher terminal velocities and are further from saturation.  As the evaporating melts approach the surface, their terminal velocities decrease and their saturation increases.  If some or all of the vapor produced by this process were to be lost from the planet with the primary atmosphere, melting and partial evaporation of mm to cm-sized pebbles in the atmosphere could impart an isotopic and elemental signature on the planet as a whole. Escape of volatilized species would eventually result from impact-driven heating and hydrodynamic escape of the H-rich atmosphere \citep{Biersteker2021}.  Detailed models of the fate of vapor deposited high in these primary atmospheres is warranted.

\subsection{Applications to objects in the terrestrial atmosphere}
Tektites are a type of glassy, silica-rich impact ejecta formed by melting and subsequent quenching of terrestrial sediments during hypervelocity meteorite impacts \citep[e.g.,][]{Howard2011}. Before quenching to glass, tektites experienced extreme heating while entrained in a hot vapor plume upon ejection from the impact site \citep[e.g.,][]{Koeberl1986, Macris2018}.  \cite{Ni2021} compared laser-heating aerodynamic levitation experiments for Cu isotope fractionation to the fractionation of Cu isotopes observed in tektites.  They derived a saturation value of $0.93\pm 0.01$ for the natural tektite data using  best fits to the data assuming Rayleigh fractionation and Equation \ref{Eqn:bigdelta}.  Ni et al. describe various potential explanations for these high apparent saturation values, including limitations due to finite diffusion in the melts, and bubble stripping in which exsolved gases produce bubbles, perhaps in equilibrium, followed by release of the equilibrated gases as the bubbles reach the surface \citep{Melosh2004}.  However, since the pressures in which the tektites are entrained are sufficiently high ($\ge 1$ bar), saturation near unity would be expected. Another possibility is that the saturation $<1$ is a result of the headwind velocity of gas over the surfaces of the projectiles while they are molten.  At face value, we can treat the tektites as spheres and use S($V_d$)/S($0$)$\sim$S with Figure  \ref{Fig:saturation} to constrain  $V_d/V_t$.  The result is a value for $V_d/V_t$ of about $0.025$.  This value may be of use in constraining the relationship between temperature of enveloping gas and differential velocity between gas and melt in the ejecta plumes.  Indeed, \cite{Magna2021} used the  concept of velocity ratios as formulated for a planar surface by  \cite{Charnoz2021} to suggest that the ratio of advective velocity to thermal velocity in tektite plume gases implies  $S > 0.98$ for potassium.  

\subsection{Potential applications to the solar protoplanetary disk}

We motivated this work in the Introduction by calling attention to the fact that chondrules exhibit limited isotope fractionation while igneous CAIs usually have significant heavy isotope enrichment for major elements, including the isotopes of Mg and Si, that are attributable to evaporation.  The explanation has been that chondrules melted in the presence of gas at near-equilibrium vapor pressures while CAIs melted in a vacuum, far from equilibrium \citep{Galy2000-1751,Young2004,Cuzzi2006-483}.   The results we present here raise the possibility that the reason igneous CAIs exhibit clear isotopic signatures of partial evaporation is in part due to a significant headwind velocity.  Similarly, a significant headwind is precluded in the case of chondrules.  However, simple equations based on the differences between the azimuthal velocities of pressure-supported gas and azimuthal velocities of dust and ``pebbles" in an active accretion disk \citep[e.g.\,][]{Armitage2019} shows that in order to reach values of  $V_d/V_t$ sufficient to lower saturation, the molten rocky bodies need to be on the order of at least a meter in size. This is perhaps not surprising  as it is another manifestation of the fact that objects with sizes $\ge 1$ meter achieve truly Keplerian orbital velocities \citep{Weidenschilling1977}, and thus maximum velocities relative to the disk gas.  In any event, the maximum headwinds encountered by mm to meter-sized objects in the disk are about $200$ m/s at 1AU from the star, and are not sufficient to be the cause of melting.  In order to apply the concept of relative gas velocities to the problem of isotope fractionation during chondrule and CAI melting, an analysis of the energetic events responsible for the melting  is required.  These events are clearly distinct from the simple orbital headwind in the disk.  Isotopic constraints on  $V_d/V_t$ may one day be useful as arbiters for various competing models for chondrule formation and CAI melting, including recent models for forming chondrules by planetesimal impacts \citep[e.g.\,][]{Lichtenberg2018}.  For example, in an impact plume, molten chondrules could not have experienced headwind velocities greater than about $0.01$ times the thermal velocities of the surrounding gas else they would have experienced fractionation of isotopes of Mg, Si, and other elements of similar or greater volatility (Figure \ref{Fig:saturation}), fractionations that are not observed.

\section{Conclusions}

We find that the isotopic fractionation associated with laser-heating aerodynamic levitation experiments is consistent with the velocity of flowing gas as the primary control on the fractionation.  The data are well explained where the gas is treated as a low-viscosity fluid that flows with high Reynolds number over the surfaces of the molten spheres with minimal drag.  The recognition that it is the ratio of flow velocity to thermal velocity that controls fractionation allows for extrapolation to other environments in which molten rock encounters gas with appreciable headwinds.  In this way, in some circumstances, the degree of isotope fractionation attending evaporation is as much a velocimeter as it is a barometer.    

\section*{Acknowledgements}
Funding was provided by NASA Emerging Worlds program grant number 80NSSC19K0511 to EDY and CM.  This work was partially supported by a Deutsche Forschungsgemeinschaft (DFG, German Research Foundation) research fellowship to QRS (project number 440227108).  QRS's work was also under the auspices of the U.S. Department of Energy by Lawrence Livermore National Laboratory under contract DE-AC52-07NA27344 with release number LLNL-JRNL-825062-DRAFT.  We thank Paolo Sossi and Andrew Davis for constructive reviews that improved the final product.



 \bibliographystyle{chicago}

\begin{thebibliography}{}

\bibitem[\protect\citeauthoryear{Alderman, Lazareva, Wilding, Benmore, Heald,
  Johnson, Johnson, Hah, Sendelbach, Tamalonis, Skinner, Parise, and
  Weber}{Alderman et~al.}{2017}]{Alderman2017}
Alderman, O., L.~Lazareva, M.~Wilding, C.~Benmore, S.~Heald, C.~Johnson,
  J.~Johnson, H.-Y. Hah, S.~Sendelbach, A.~Tamalonis, L.~Skinner, J.~Parise,
  and J.~Weber (2017).
\newblock Local structural variation with oxygen fugacity in {Fe$_2$SiO$_4+x$}
  fayalitic iron silicate melts.
\newblock {\em Geochimica et Cosmochimica Acta\/}~{\em 203}, 15--36.

\bibitem[\protect\citeauthoryear{All\'{e}gre, {Manh\'{e}s}, and
  {Lewin}}{All\'{e}gre et~al.}{2001}]{Allegre2001}
All\'{e}gre, C., G.~{Manh\'{e}s}, and {\'E}.~{Lewin} (2001).
\newblock Chemical composition of the earth and the volatility control on
  planetary genetics.
\newblock {\em Earth and Planetary Science Letters\/}, 49--69.

\bibitem[\protect\citeauthoryear{Armitage}{Armitage}{2019}]{Armitage2019}
Armitage, P.~J. (2019).
\newblock {\em Physical Processes in Protoplanetary Disks}, pp.\  1--150.
\newblock Berlin, Heidelberg: Springer Berlin Heidelberg.

\bibitem[\protect\citeauthoryear{Badro, Sossi, Deng, Borensztajn, Wehr, and
  Ryerson}{Badro et~al.}{2021}]{Badro2021}
Badro, J., P.~A. Sossi, Z.~Deng, S.~Borensztajn, N.~Wehr, and F.~J. Ryerson
  (2021).
\newblock Experimental investigation of elemental and isotopic evaporation
  processes by laser heating in an aerodynamic levitation furnace.
\newblock {\em Comptes Rendus. G\'eoscience\/}~{\em 353\/}(1), 101--114.

\bibitem[\protect\citeauthoryear{Biersteker and Schlichting}{Biersteker and
  Schlichting}{2020}]{Biersteker2021}
Biersteker, J.~B. and H.~E. Schlichting (2020).
\newblock {Losing oceans: The effects of composition on the thermal component
  of impact-driven atmospheric loss}.
\newblock {\em Monthly Notices of the Royal Astronomical Society\/}~{\em
  501\/}(1), 587--595.

\bibitem[\protect\citeauthoryear{Braun, Sousa, and Paniagua}{Braun
  et~al.}{2018}]{Braun2018}
Braun, J., J.~Sousa, and G.~Paniagua (2018).
\newblock Numerical assessment of the convective heat transfer in rotating
  detonation combustors using a reduced-order model.
\newblock {\em Applied Sciences\/}~{\em 8\/}(6).

\bibitem[\protect\citeauthoryear{Brouwers, Ormel, Bonsor, and Vazan}{Brouwers
  et~al.}{2021}]{Brouwers_2021}
Brouwers, M., C.~W. Ormel, A.~Bonsor, and A.~Vazan (2021).
\newblock How planets grow by pebble accretion.
\newblock {\em Astronomy \& Astrophysics\/}~{\em in press}.

\bibitem[\protect\citeauthoryear{Charnoz, Sossi, Lee, Siebert, Hyodo, Allibert,
  Pignatale, Landeau, Oza, and Moynier}{Charnoz et~al.}{2021}]{Charnoz2021}
Charnoz, S., P.~A. Sossi, Y.-N. Lee, J.~Siebert, R.~Hyodo, L.~Allibert, F.~C.
  Pignatale, M.~Landeau, A.~V. Oza, and F.~Moynier (2021).
\newblock Tidal pull of the {Earth strips the proto-Moon} of its volatiles.
\newblock {\em Icarus\/}~{\em 364}, 114451.

\bibitem[\protect\citeauthoryear{Chayes}{Chayes}{1971}]{Chayes1971}
Chayes, F. (1971).
\newblock {\em Ratio Correlation: A Manual for Students of Petrology and
  Geochemistry}.
\newblock University of Chicago Press.

\bibitem[\protect\citeauthoryear{Costa, Jacobson, and {Fegley Jr.}}{Costa
  et~al.}{2017}]{Costa2017}
Costa, G.~C., N.~S. Jacobson, and B.~{Fegley Jr.} (2017).
\newblock Vaporization and thermodynamics of forsterite-rich olivine and some
  implications for silicate atmospheres of hot rocky exoplanets.
\newblock {\em Icarus\/}~{\em 289}, 42--55.

\bibitem[\protect\citeauthoryear{Cuzzi and Alexander}{Cuzzi and
  Alexander}{2006}]{Cuzzi2006-483}
Cuzzi, J.~N. and C.~M. O.~D. Alexander (2006).
\newblock Chondrule formation in particle-rich nebular regions at least
  hundreds of kilometers across.
\newblock {\em Nature\/}~{\em 441}, 483--485.

\bibitem[\protect\citeauthoryear{{Davis, A.M. }, Hashimoto, {Clayton, R.N.},
  and {Mayeda, T.K.}}{{Davis, A.M. } et~al.}{1990}]{Davis1990}
{Davis, A.M. }, A.~Hashimoto, {Clayton, R.N.}, and {Mayeda, T.K.} (1990).
\newblock Isotope mass fractionation during evaporation of{ Mg$_2$SiO$_4$}.
\newblock {\em Nature\/}~{\em 347}, 655--658.

\bibitem[\protect\citeauthoryear{Depredurand, Castanet, and
  Lemoine}{Depredurand et~al.}{2010}]{Depredurand2010}
Depredurand, V., G.~Castanet, and F.~Lemoine (2010).
\newblock Heat and mass transfer in evaporating droplets in interaction:
  Influence of the fuel.
\newblock {\em International Journal of Heat and Mass Transfer\/}~{\em
  53\/}(17), 3495--3502.

\bibitem[\protect\citeauthoryear{Eisenklam, Arunachalam, and Weston}{Eisenklam
  et~al.}{1967}]{Eisenklam1967}
Eisenklam, P., S.~Arunachalam, and J.~Weston (1967).
\newblock Evaporation rates and drag resistance of burning drops.
\newblock {\em Symposium (International) on Combustion\/}~{\em 11\/}(1),
  715--728.

\bibitem[\protect\citeauthoryear{{Fedkin}, {Grossman}, and {Ghiorso}}{{Fedkin}
  et~al.}{2006}]{Fedkin2006}
{Fedkin}, A.~V., L.~{Grossman}, and M.~S. {Ghiorso} (2006).
\newblock {Vapor pressures and evaporation coefficients for melts of
  ferromagnesian chondrule-like compositions}.
\newblock {\em Geochimica et Cosmochimica Acta\/}~{\em 70\/}(1), 206--223.

\bibitem[\protect\citeauthoryear{{Fegley} and {Cameron}}{{Fegley} and
  {Cameron}}{1987}]{Fegley1987}
{Fegley}, B.~J. and A.~G.~W. {Cameron} (1987).
\newblock A vaporization model for iron/silicate fractionation in the mercury
  protoplanet.
\newblock {\em Earth and Planetary Science Letters\/}~{\em 82}, 207--222.

\bibitem[\protect\citeauthoryear{Feng and Michaelides}{Feng and
  Michaelides}{2001}]{Feng2001}
Feng, Z.-G. and E.~E. Michaelides (2001).
\newblock {Drag Coefficients of Viscous Spheres at Intermediate and High
  Reynolds Numbers }.
\newblock {\em Journal of Fluids Engineering\/}~{\em 123\/}(4), 841--849.

\bibitem[\protect\citeauthoryear{{Fressin}, {Torres}, {Charbonneau}, {Bryson},
  {Christiansen}, {Dressing}, {Jenkins}, {Walkowicz}, and {Batalha}}{{Fressin}
  et~al.}{2013}]{Fressin2013a}
{Fressin}, F., G.~{Torres}, D.~{Charbonneau}, S.~T. {Bryson},
  J.~{Christiansen}, C.~D. {Dressing}, J.~M. {Jenkins}, L.~M. {Walkowicz}, and
  N.~M. {Batalha} (2013).
\newblock {The false positive rate of Kepler and the occurrence of planets}.
\newblock {\em The Astrophysical Journal\/}~{\em 766}, 81.

\bibitem[\protect\citeauthoryear{Frost}{Frost}{2018}]{Frost2018}
Frost, B.~R. (2018).
\newblock {\em Chapter 1.INTRODUCTION TO OXYGEN FUGACITY AND ITS PETROLOGIC
  IMPORTANCE}, pp.\  1--10.
\newblock De Gruyter.

\bibitem[\protect\citeauthoryear{Galy, Young, Ash, and O'Nions}{Galy
  et~al.}{2000}]{Galy2000-1751}
Galy, A., E.~D. Young, R.~D. Ash, and R.~K. O'Nions (2000).
\newblock The formation of chondrules at high gas pressures in the solar
  nebula.
\newblock {\em Science\/}~{\em 290}, 1751--1753.

\bibitem[\protect\citeauthoryear{{Ghiorso}, {Hirschmann}, {Reiners}, and
  {Kress}}{{Ghiorso} et~al.}{2002}]{Ghiorso2002}
{Ghiorso}, M.~S., M.~M. {Hirschmann}, P.~W. {Reiners}, and V.~C. {Kress}
  (2002).
\newblock The pmelts: A revision of melts for improved calculation of phase
  relations and major element partitioning related to partial melting of the
  mantle to 3 gpa.
\newblock {\em Geochemistry, Geophysics, Geosystems\/}~{\em 3}, 1030.

\bibitem[\protect\citeauthoryear{{Ginzburg}, {Schlichting}, and
  {Sari}}{{Ginzburg} et~al.}{2016}]{Ginzburg2016a}
{Ginzburg}, S., H.~E. {Schlichting}, and R.~{Sari} (2016).
\newblock {Super-Earth atmospheres: Self-consistent gas accretion and
  retention}.
\newblock {\em The Astrophysical Journal\/}~{\em 825}, 29.

\bibitem[\protect\citeauthoryear{{Gupta} and {Schlichting}}{{Gupta} and
  {Schlichting}}{2019}]{Gupta2019a}
{Gupta}, A. and H.~E. {Schlichting} (2019).
\newblock {Sculpting the valley in the radius distribution of small exoplanets
  as a by-product of planet formation: the core-powered mass-loss mechanism}.
\newblock {\em Monthly Notices of the Royal Astronomical Society\/}~{\em
  487\/}(1), 24--33.

\bibitem[\protect\citeauthoryear{Halliday and Porcelli}{Halliday and
  Porcelli}{2001}]{Halliday2001}
Halliday, A.~N. and D.~Porcelli (2001).
\newblock In search of lost planets - the paleocosmochemistry of the inner
  solar system.
\newblock {\em Earth and Planetary Science Letters\/}~{\em 192}, 545--559.

\bibitem[\protect\citeauthoryear{Hastie and Bonnell}{Hastie and
  Bonnell}{1986}]{Hastie1986}
Hastie, J. and D.~Bonnell (1986).
\newblock {A predictive thermodynamic model of oxide and halide glass phase
  equilibria}.
\newblock {\em Journal of Non-Crystalline Solids\/}~{\em 84\/}(1), 151--158.

\bibitem[\protect\citeauthoryear{Howard}{Howard}{2011}]{Howard2011}
Howard, K.~T. (2011).
\newblock Volatile enhanced dispersal of high velocity impact melts and the
  origin of tektites.
\newblock {\em Proceedings of the Geologists' Association\/}~{\em 122\/}(3),
  363--382.

\bibitem[\protect\citeauthoryear{Jordan, Tang, Kohl, and Young}{Jordan
  et~al.}{2019}]{Jordan2019}
Jordan, M.~K., H.~Tang, I.~E. Kohl, and E.~D. Young (2019).
\newblock Iron isotope constraints on planetesimal core formation in the early
  solar system.
\newblock {\em Geochimica et Cosmochimica Acta\/}~{\em 246}, 461--477.

\bibitem[\protect\citeauthoryear{Koeberl}{Koeberl}{1986}]{Koeberl1986}
Koeberl, C. (1986).
\newblock {Muong Nong} type tektites from the moldavite and north american
  strewn fields?
\newblock {\em Journal of Geophysical Research: Planets\/}~{\em 91\/}(B13),
  E253.

\bibitem[\protect\citeauthoryear{Kreutzberger, Drake, and Jones}{Kreutzberger
  et~al.}{1986}]{Kreutzberger1986}
Kreutzberger, M.~E., M.~J. Drake, and J.~H. Jones (1986).
\newblock Origin of the {Earth's Moon}: Constraints from alkali volatile trace
  elements.
\newblock {\em Geochemica et Cosmochimica Acta\/}~{\em 50}, 91--98.

\bibitem[\protect\citeauthoryear{Kurose, Makino, Komori, Nakamura, Akamatsu,
  and Katsuki}{Kurose et~al.}{2003}]{Kurose2003}
Kurose, R., H.~Makino, S.~Komori, M.~Nakamura, F.~Akamatsu, and M.~Katsuki
  (2003).
\newblock Effects of outflow from the surface of a sphere on drag, shear lift,
  and scalar diffusion.
\newblock {\em Physics of Fluids\/}~{\em 15\/}(8), 2338--2351.

\bibitem[\protect\citeauthoryear{{Lee} and {Chiang}}{{Lee} and
  {Chiang}}{2016}]{Lee2016a}
{Lee}, E.~J. and E.~{Chiang} (2016).
\newblock {Breeding Super-Earths and Birthing Super-puffs in Transitional
  Disks}.
\newblock {\em The Astrophysical Journal\/}~{\em 817}, 90.

\bibitem[\protect\citeauthoryear{Lemmon and Jacobsen}{Lemmon and
  Jacobsen}{2004}]{Lemmon2004}
Lemmon, E.~W. and R.~T. Jacobsen (2004).
\newblock Viscosity and thermal conductivity equations for nitrogen, oxygen,
  argon, and air.
\newblock {\em International Journal of Thermophysics\/}~{\em 25\/}(1), 21--69.

\bibitem[\protect\citeauthoryear{Lichtenberg, Golabek, Dullemond,
  {Sch\"{o}nb\"{a}chler}, Gerya, and Meyer}{Lichtenberg
  et~al.}{2018}]{Lichtenberg2018}
Lichtenberg, T., G.~J. Golabek, C.~P. Dullemond, M.~{Sch\"{o}nb\"{a}chler},
  T.~V. Gerya, and M.~R. Meyer (2018).
\newblock Impact splash chondrule formation during planetesimal recycling.
\newblock {\em Icarus\/}~{\em 302}, 27--43.

\bibitem[\protect\citeauthoryear{Macris, Asimow, Badro, Eiler, Zhang, and
  Stolper}{Macris et~al.}{2018}]{Macris2018}
Macris, C.~A., P.~D. Asimow, J.~Badro, J.~M. Eiler, Y.~Zhang, and E.~M. Stolper
  (2018).
\newblock Seconds after impact: Insights into the thermal history of impact
  ejecta from diffusion between lechatelierite and host glass in tektites and
  experiments.
\newblock {\em Geochimica et Cosmochimica Acta\/}~{\em 241}, 69--94.

\bibitem[\protect\citeauthoryear{Magna, Jiang, Sk\'{a}la, Wang, Sossi, and
  \v{Z}\'{a}k}{Magna et~al.}{2021}]{Magna2021}
Magna, T., Y.~Jiang, R.~Sk\'{a}la, K.~Wang, P.~A. Sossi, and K.~\v{Z}\'{a}k
  (2021).
\newblock Potassium elemental and isotope constraints on the formation of
  tektites and element loss during impacts.
\newblock {\em Geochimica et Cosmochimica Acta\/}~{\em 312}, 321--342.

\bibitem[\protect\citeauthoryear{McCoy, Dickinson, and Lofgren}{McCoy
  et~al.}{1999}]{McCoy1999}
McCoy, T.~J., T.~L. Dickinson, and G.~E. Lofgren (1999).
\newblock {Partial melting of the Indarch (EH4) Meteorite: A textural, chemical
  and phase relations view of melting and melt migration}.
\newblock {\em Meteoritics and Planetary Science\/}~{\em 34\/}(5), 735--746.

\bibitem[\protect\citeauthoryear{McDonald}{McDonald}{2015}]{McDonald2015}
McDonald, K.~T. (2015).
\newblock Pressure in fluid flow past a spheere.
\newblock {\em Course Notes, Princeton University\/}~{\em 347}, 6.

\bibitem[\protect\citeauthoryear{{Melosh} and {Artemieva}}{{Melosh} and
  {Artemieva}}{2004}]{Melosh2004}
{Melosh}, H.~J. and N.~{Artemieva} (2004).
\newblock {How Does Tektite Glass Lose Its Water?}
\newblock In S.~{Mackwell} and E.~{Stansbery} (Eds.), {\em Lunar and Planetary
  Science Conference}, Lunar and Planetary Science Conference, pp.\  1723.

\bibitem[\protect\citeauthoryear{Moore}{Moore}{1963}]{Moore1963}
Moore, D.~W. (1963).
\newblock The boundary layer on a spherical gas bubble.
\newblock {\em Journal of Fluid Mechanics\/}~{\em 16\/}(2), 161--176.

\bibitem[\protect\citeauthoryear{Nagahara and Ozawa}{Nagahara and
  Ozawa}{1996}]{Nagahara1996-1445}
Nagahara, H. and K.~Ozawa (1996).
\newblock Evaporation of forsterite in h2 gas.
\newblock {\em Geochimica et Cosmochimica Acta\/}~{\em 60}, 1445--1459.

\bibitem[\protect\citeauthoryear{{Ni}, {Macris}, {Darling}, and {Shahar}}{{Ni}
  et~al.}{2021}]{Ni2021}
{Ni}, P., C.~A. {Macris}, E.~A. {Darling}, and A.~{Shahar} (2021).
\newblock {Evaporation-induced copper isotope fractionation: Insights from
  laser levitation experiments}.
\newblock {\em Geochimica et Cosmochimica Acta\/}~{\em 298}, 131--148.

\bibitem[\protect\citeauthoryear{Ormel}{Ormel}{2017}]{Ormel2017}
Ormel, C.~W. (2017).
\newblock {\em {The Emerging Paradigm of Pebble Accretion}}, Volume 445, pp.\
  197.

\bibitem[\protect\citeauthoryear{Palme and Boynton}{Palme and
  Boynton}{1993}]{Palme1993-979}
Palme, H. and W.~Boynton (1993).
\newblock Meteoritic constraints on conditions in the solar nebula.
\newblock In E.~Levy and J.~Lunine (Eds.), {\em Protostars and Planets III},
  pp.\  979--1004. Tucson: University of Arizona Press.

\bibitem[\protect\citeauthoryear{Paniello, Day, and Moynier}{Paniello
  et~al.}{2012}]{Paniello2012}
Paniello, R.~C., J.~M.~D. Day, and F.~Moynier (2012).
\newblock Zinc isotopic evidence for the origin of the moon.
\newblock {\em Nature\/}~{\em 490}, 376--379.

\bibitem[\protect\citeauthoryear{Polyakov and Mineev}{Polyakov and
  Mineev}{2000}]{Polyakov2000-849}
Polyakov, V. and S.~Mineev (2000).
\newblock The use of {M\"{o}ssbauer} spectroscopy in stable isotope
  geochemistry.
\newblock {\em Geochimica et Cosmochimica Acta\/}~{\em 64}, 849--865.

\bibitem[\protect\citeauthoryear{Richter, Davis, Ebel, and Hashimoto}{Richter
  et~al.}{2002}]{Richter2002-521}
Richter, F.~M., A.~M. Davis, D.~S. Ebel, and A.~Hashimoto (2002).
\newblock Elemental and isotopic fractionation of type b calcium- aluminum-rich
  inclusions: Experiments, theoretical considerations, and constraints on their
  thermal evolution.
\newblock {\em Geochimica et Cosmochimica Acta\/}~{\em 66\/}(3), 521--540.

\bibitem[\protect\citeauthoryear{{Richter, F.M.}, {Janney, P.E. }and
  {Mendybaev, R.A.}, {Davis, A.M.}, and Wadhwa}{{Richter, F.M.}
  et~al.}{2007}]{Richter2007}
{Richter, F.M.}, {Janney, P.E. }and {Mendybaev, R.A.}, {Davis, A.M.}, and
  M.~Wadhwa (2007).
\newblock Elemental and isotopic fractionatiuon of type b cai-like liquids by
  evaporation.
\newblock {\em Geochimica et Cosmochimica Acta\/}~{\em 71}, 5544--5564.

\bibitem[\protect\citeauthoryear{Salazar and S\'anchez-Lavega}{Salazar and
  S\'anchez-Lavega}{1999}]{Salazar1999}
Salazar, A. and A.~S\'anchez-Lavega (1999).
\newblock Low temperature thermal diffusivity measurements of gases by the
  mirage technique.
\newblock {\em Review of Scientific Instruments\/}~{\em 70\/}(1), 98--103.

\bibitem[\protect\citeauthoryear{{Schaefer} and Fegley}{{Schaefer} and
  Fegley}{2004}]{Schaefer2004}
{Schaefer}, L. and B.~Fegley (2004).
\newblock {A thermodynamic model of high temperature lava vaporization on Io}.
\newblock {\em Icarus\/}~{\em 169\/}(1), 216--241.

\bibitem[\protect\citeauthoryear{{Schaefer} and {Fegley}}{{Schaefer} and
  {Fegley}}{2009}]{Schaefer2009}
{Schaefer}, L. and B.~{Fegley} (2009).
\newblock {Chemistry of Silicate Atmospheres of Evaporating Super-Earths}.
\newblock {\em The Astrophysical Journal Letters\/}~{\em 703\/}(2), L113--L117.

\bibitem[\protect\citeauthoryear{Schauble}{Schauble}{2011}]{Schauble2011}
Schauble, E.~A. (2011).
\newblock First-principles estimates of equilibrium magnesium isotope
  fractionation in silicate, oxide, and carbonate and hexaaquamagnesium(2+)
  crystals.
\newblock {\em Geochimica et Cosmochimica Acta\/}~{\em 75}, 844--869.

\bibitem[\protect\citeauthoryear{Shahar and Young}{Shahar and
  Young}{2007}]{Shahar2007}
Shahar, A. and E.~D. Young (2007).
\newblock Astrophysics of cai formation as revealed by silicon isotope
  la-mc-icpms of an igneous cai.
\newblock {\em Earth and Planetary Science Letters\/}~{\em 257}, 497--510.

\bibitem[\protect\citeauthoryear{Sossi, Klemme, O'Neill, Berndt, and
  Moynier}{Sossi et~al.}{2019}]{Sossi2019}
Sossi, P.~A., S.~Klemme, H.~S. O'Neill, J.~Berndt, and F.~Moynier (2019).
\newblock Evaporation of moderately volatile elements from silicate melts:
  experiments and theory.
\newblock {\em Geochimica et Cosmochimica Acta\/}~{\em 260}, 204--231.

\bibitem[\protect\citeauthoryear{Tang, Szumila, Trail, and Young}{Tang
  et~al.}{2021}]{Tang2021}
Tang, H., I.~Szumila, D.~Trail, and E.~D. Young (2021).
\newblock Experimental determination of the effect of cr on mg isotope
  fractionation between spinel and forsterite.
\newblock {\em Geochimica et Cosmochimica Acta\/}~{\em 296}, 152--169.

\bibitem[\protect\citeauthoryear{Weber, Felten, and Nordine}{Weber
  et~al.}{1996}]{Weber1996}
Weber, J.~K., J.~J. Felten, and P.~C. Nordine (1996).
\newblock Laser hearth melt processing of ceramic materials.
\newblock {\em Review of Scientific Instruments\/}~{\em 67\/}(2), 522--524.

\bibitem[\protect\citeauthoryear{{Weidenschilling}}{{Weidenschilling}}{1977}]{Weidenschilling1977}
{Weidenschilling}, S.~J. (1977).
\newblock {Aerodynamics of solid bodies in the solar nebula.}
\newblock {\em Monthly Notices of the Royal Astronomical Society\/}~{\em 180},
  57--70.

\bibitem[\protect\citeauthoryear{Wimpenny, Marks, Knight, Rolison, Borg,
  Eppich, Badro, Ryerson, Sanborn, Huyskens, and Yin}{Wimpenny
  et~al.}{2019}]{Wimpenny2019}
Wimpenny, J., N.~Marks, K.~Knight, J.~M. Rolison, L.~Borg, G.~Eppich, J.~Badro,
  F.~J. Ryerson, M.~Sanborn, M.~H. Huyskens, and Q.-z. Yin (2019).
\newblock Experimental determination of zn isotope fractionation during
  evaporative loss at extreme temperatures.
\newblock {\em Geochimica et Cosmochimica Acta\/}~{\em 259}, 391--411.

\bibitem[\protect\citeauthoryear{Wombacher, Eisenhauer, Heuser, and
  Weyer}{Wombacher et~al.}{2009}]{Wombacher2009}
Wombacher, F., A.~Eisenhauer, A.~Heuser, and S.~Weyer (2009).
\newblock Separation of mg{,} ca and fe from geological reference materials for
  stable isotope ratio analyses by mc-icp-ms and double-spike tims.
\newblock {\em J. Anal. At. Spectrom.\/}~{\em 24}, 627--636.

\bibitem[\protect\citeauthoryear{Young, Tonui, Manning, Schauble, and
  Macris}{Young et~al.}{2009}]{Young2009}
Young, E., E.~Tonui, C.~Manning, E.~Schauble, and C.~Macris (2009).
\newblock Spinel-olivine magnesium isotope thermometry in the mantle and
  implications for the mg isotopic composition of earth.
\newblock {\em Earth and Planetary Science Letters\/}~{\em 288\/}(3), 524--533.

\bibitem[\protect\citeauthoryear{Young}{Young}{2000}]{Young2000-321}
Young, E.~D. (2000).
\newblock Assessing the implications of k isotope cosmochemistry for
  evaporation in the preplanetary solar nebula.
\newblock {\em Earth and Planetary Science Letters\/}~{\em 183\/}(1-2),
  321--333.
\newblock Times Cited: 9.

\bibitem[\protect\citeauthoryear{Young and Galy}{Young and
  Galy}{2004}]{Young2004}
Young, E.~D. and A.~Galy (2004).
\newblock The isotope geochemistry and cosmochemistry of magnesium.
\newblock In {\em Geochemistry of Non-Traditional Stable Isotopes}, Volume~55
  of {\em Reviews in Mineralogy \& Geochemistry}, pp.\  197--230. MSA.
\newblock Times Cited: 1.

\bibitem[\protect\citeauthoryear{{Young}, {Shahar}, {Nimmo}, {Schlichting},
  {Schauble}, {Tang}, and {Labidi}}{{Young} et~al.}{2019}]{Young2019}
{Young}, E.~D., A.~{Shahar}, F.~{Nimmo}, H.~E. {Schlichting}, E.~A. {Schauble},
  H.~{Tang}, and J.~{Labidi} (2019).
\newblock Near-equilibrium isotope fractionation during planetesimal
  evaporation.
\newblock {\em Icarus\/}~{\em 323}, 1--15.

\end{thebibliography}

\section*{Appendix 1: Methods}
\subsection*{Experiments}  
The starting material for our experiments was a simplified, synthetic metal-free enstatite chondrite powder (similar to Indarch, EH4)  made by mixing reagent grade oxide powders under ultrapure isopropanol in an agate mortar and pestle.  The powder is composed of 41.83\% SiO$_2$, 29.12\% FeO, 22.61\% MgO, 2.67\% Cr$_2$O$_3$, 2.16\% Al$_2$O$_3$, 0.82\% Na$_2$O, 0.69\% CaO, and 0.09\% K$_2$O by weight.  By adding Fe in oxidized form, the starting material is inherently more oxidized than enstatite chondrites.  Vaporization experiments were performed in the High-Temperature Conical Nozzle Levitation (HT-CNL) System (this apparatus is sometimes also referred to as an aerodynamic levitation laser furnace) at the High Pressure and Temperature Geochemistry lab at Indiana University-Purdue University Indianapolis (IUPUI).  The apparatus components and mode of operation  are described in \cite{Ni2021} (Figure \ref{Fig:apparatus}).  

\begin{figure}
\centering
\captionsetup{width=0.75\linewidth}
 \includegraphics[width=3.7in]{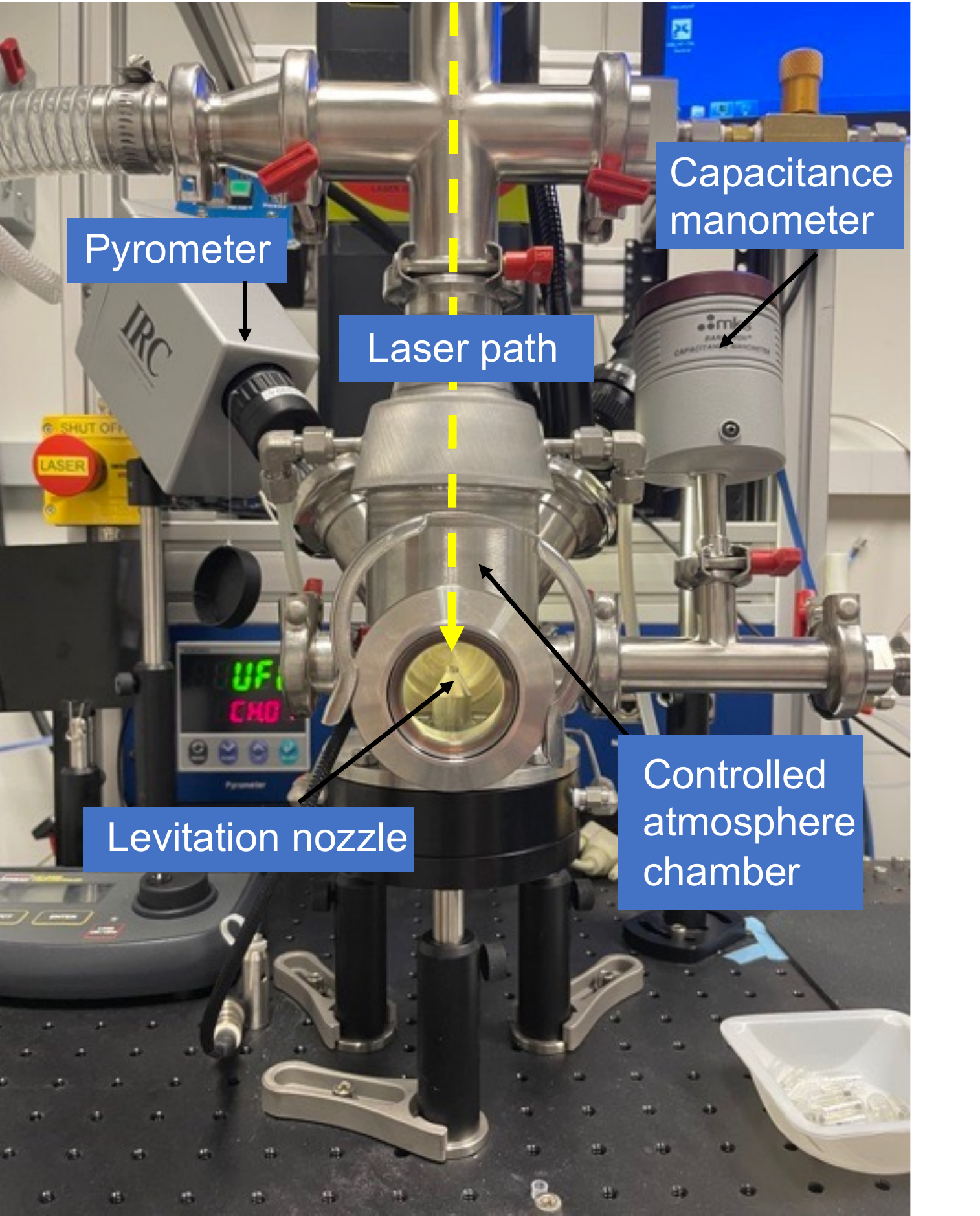}
\caption{Aerodynamic levitation laser-heating system at IUPUI.  The nozzle and position of the levitated spheres can be seen through the window in the center of the image.  Pressures are measured at the capacitance manometer on the right.  The port for pyrometry feedback is at the left.} 
\label{Fig:apparatus} 
\end{figure}

In these experiments, a spherical sample, $\sim 2$ mm in diameter, is levitated by a gas (or gas mixture) issuing from a conical nozzle while being heated with a 40 W CO$_2$ IR laser ($10.6 \mu$m wavelength). Sample temperature is measured by a pyrometer that provides feedback for control of the laser power to achieve the desired experimental temperature-time (T-t) conditions. Heating and levitation take place in a controlled-atmosphere chamber, with gas mixing capabilities. Prior to levitation, $\sim 15$ mg aliquots of the synthetic chondrite powder were fused into spheres suitable for levitation in a water-cooled oxygen-free hearth plate by defocused laser heating. This method of sample preparation results in negligible sample contamination and material loss \citep{Weber1996}, including in Fe-bearing systems \citep{Alderman2017}. After weighing, pre-fused spherical  samples were heated to 2273 or 2323 K  for 120 to 480 s while levitated in either 99.999\% purity N$_2$, or a mixture of $95\%$ Ar , $\sim 4.5\%$ CO and $\sim 0.5\%$ CO$_2$.  The oxygen fugacity in N$_2$ gas was not controlled externally.  The Ar-CO-CO$_2$ mixture imposed an oxygen fugacity relative to the iron-w\"{u}stite equilibrium value, $\log{f_\mathrm{O_2}} -\log{f_\mathrm{O_2}}({\rm IW})$, of $-0.5$ \citep[e.g.,][]{Frost2018}.  Samples were quenched to glass by cutting power to the laser.  A sample temperature vs.\ time plot is shown in Figure \ref{Fig:temp_vs_time}.  

\begin{figure}
\centering
\captionsetup{width=0.7\linewidth}
\includegraphics[width=3.8in]{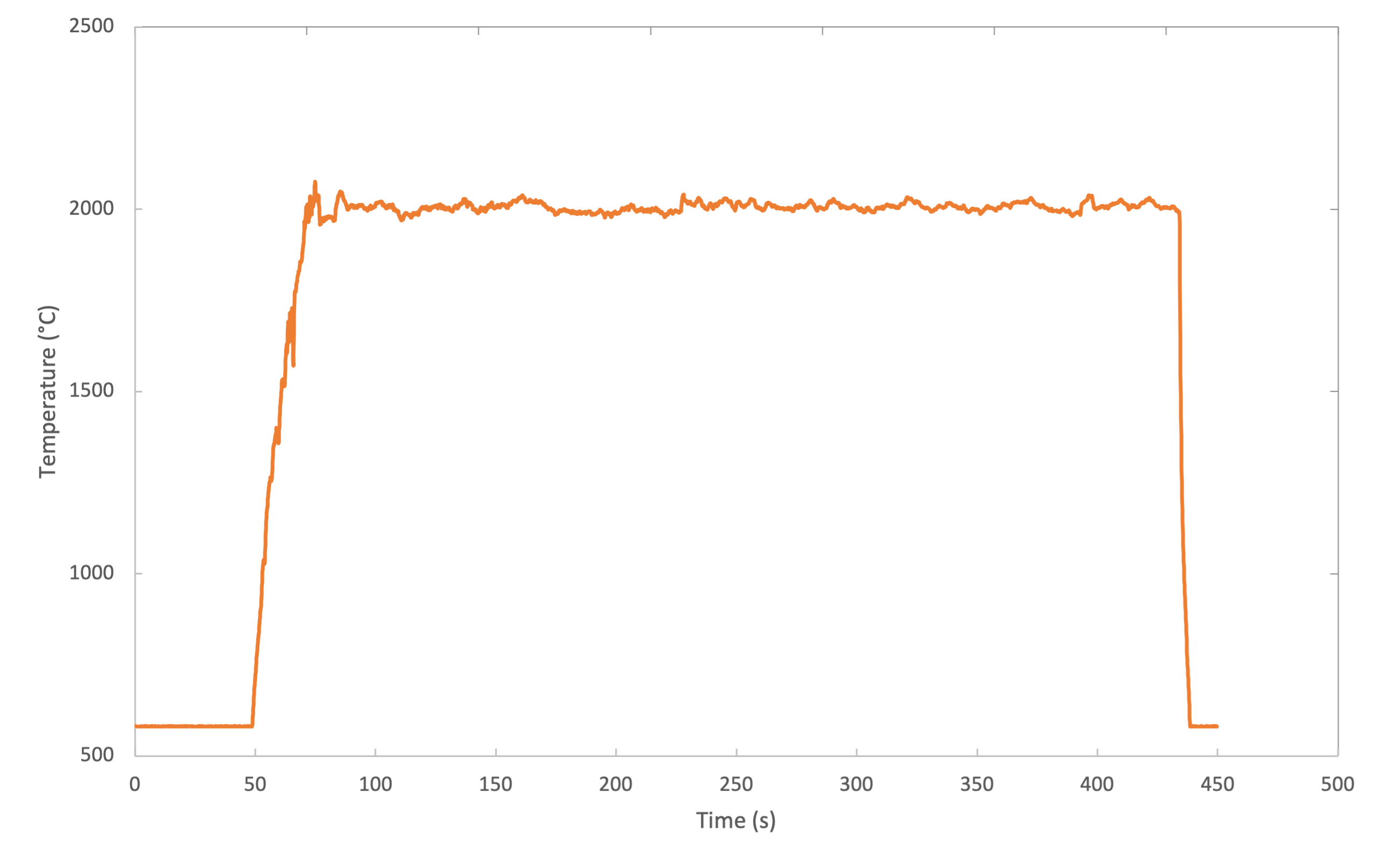}
\caption{Temperature vs.\ time plot for experiment EL-2.77.}
\label{Fig:temp_vs_time}
\end{figure}

The aerodynamic levitation chamber was modified for this study to allow operation at variable pressures within a limited range ($\sim 0.1$ to $1.1$ bar, depending on gas flow requirements for sample levitation). The modification involved adding a tee above the nozzle, which is connected to the levitation gas (to provide cross-flow) on one end and a vacuum hose on the other (Figure \ref{Fig:apparatus}). The vacuum hose leads to a vacuum pump with a variable conductance valve, both using an external power source. These new components draw gas and other particles out of the chamber to achieve a desired far-field pressure. A capacitance manometer interfaces with the LabView software to conduct experiments at a user-indicated controlled chamber pressure.  Experiments in this study were conducted at two ambient chamber pressures: $1.013$ bar and $0.333$ bar.   A levitation gas flow rate of 360 cc/min was used in most runs.

\subsection*{Isotopic and elemental analyses}

Experimental products were crushed into pieces and digested at the Department of Earth, Planetary and Space Sciences at University of California-Los Angeles (UCLA). In preparation for isotopic analyses using a Multi-Collector Inductively Coupled Plasma Mass Spectrometer (MC-ICPMS), Fe and Mg were extracted and purified from the melt residues using ion exchange chromatography in a Class 100 clean wet chemistry laboratory. The samples were digested in a 2:1 mixture of Omnitrace HF and HNO$_{3}$ at temperature of ~125 $^o$C for 72-96 hrs on a hot plate. The partially dissolved samples were evaporated to dryness at 125 $^o$C, redissolved in aqua regia (HCl:HNO$_{3}$ = 2:1), and placed on a hot plate at 120 $^o$C for 24-48 hrs. The sample solutions were then dried down and redissolved in 6 N HCl in preparation for Fe and Mg purification as well as elemental analysis.

The purification procedure for Fe followed \cite{Jordan2019}. 1 mL of pre-cleaned wet Bio-Rad AG1-X8 anion exchange resin in a 200-400 mesh was added to the column. The resin was then cleaned in the column using an 18 M$\Omega$ water, 1 N HNO$_3$, and 0.4 N HCl  to remove any pre-existing Fe. The column was conditioned with 5 mL of 6 N HCl, and $\sim$25-50\% of the sample solutions were loaded into the columns in 0.5-1.0 mL of 6 N HCl. To elute non-iron elements, 8 mL of 6 N HCl was used. The 6N cut was preserved in a pre-cleaned Teflon beaker for further purification of Mg. The remaining Fe was then eluted using 8 mL of 0.4 N HCl. This two-step elution was repeated again, to obtain Fe purity. The Fe cut was evaporated to dryness in Sallivex and redissolved in 2\% HNO$_{3}$ for MC-ICPMS analysis. The preserved 6 N sample cuts containing Mg and other matrix elements were dried down and redissolved in $\sim$10 N HCl for Mg purification.

A two-column technique was used for the ion-exchange chromatography purification of Mg. The purification procedure was modified from \cite{Young2009}, \cite{Wombacher2009}, and \cite{Tang2021}. We used BioRad 10 mL columns filled with 2 mL of AG50W-X8 resin in 200-400 mesh hydrogen form. The first column is to separate Ca from Mg to avoid any interference from $^{48}$Ca$^{++}$ on $^{24}$Mg$^{+}$. The second column  is used to remove Cr, alkali elements (Na and K), and Al.  Before loading the samples, the resin was pre-cleaned with  10 mL of Milli-Q water, 10 mL of 4 N HCl, 10 mL of Milli-Q water, and 10 mL of 4 N HCl, followed by conditioning with 6 mL of ~10 N HCl. Approximately 0.5 mL sample volumes in $\sim$10 N HCl were loaded on the first column.  The samples comprised ~30-50 $\mu$g of Mg. Magnesium was collected with 5 mL of ~10 N HCl, leaving Ca on the resin. Approximately 99.75\% of Ca is eliminated from Mg in this procedure. The Mg in $\sim$10 N HCl was then evaporated on a hot plate at 120 $^o$C and redissolved in 0.5-1 mL 0.5 N HCl for the second column procedure. BioRad 10 mL columns filled with 2 mL of AG50W-X8 resin in 200-400 mesh hydrogen form were used for the second purification step. The resin was washed with 10 mL of Milli-Q water, 10 mL of 0.5 N HCl, 15 mL of 6 N HCl, 10 mL of Milli-Q water, and 15 mL of 6 N HCl, followed by conditioning with 10 mL of 0.5 N HCl. After loading 0.5 to 1 mL sample solutions in 0.5 N HCl, Cr and alkali elements (Na and K) were eluted with 34 mL of 0.5 N HCl.  Aluminum was removed with 7 mL of 0.15 N HF. The columns were then washed with 2 mL Milli-Q water. Mg was eluted  with 12 mL of 2 N HCl. After purification, the Mg sample solutions were dried down and then redissolved in 2\% HNO$_{3}$ for isotopic analysis. The yields of Mg after the entire purification procedure are $>$99\%. The Mg blank is $\sim$30-40 ng, negligible relative to the amount of Mg processed through the column procedure. To monitor the accuracy of the the analytical procedures, two well-characterized geostandards, BHVO-2 and DTS-02, were digested, chemically purified, and analyzed simultaneously with the exchanged samples.

The isotopic measurements were conducted at UCLA using a ThermoFinnigan Neptune MC-ICPMS instrument. Samples were analyzed using wet plasma with a quartz dual cyclonic spray chamber. The uptake flow rate was 50 $\mu$L/min. For Fe isotopic analyses, faraday cups L2, L1, C, H1, H2, and H3 with amplifier resistors of 10$^{11}$ $\Omega$ were used to measure singly-charged ion beams of $^{53}$Cr, $^{54}$Fe, $^{56}$Fe, $^{57}$Fe, $^{58}$Fe, and $^{60}$Ni respectively. Samples were run at a mass resolving power ($m/\delta m$) of  $>$8500 to eliminate interferences from $^{40}$Ar$^{14}$N$^{+}$ and $^{40}$Ar$^{16}$O$^{+}$. For Mg isotopic analyses, faraday cups L3, L2, C, H2 and H4 with amplifier resistors of 10$^{11}$ $\Omega$   were used to measure $^{23}$Na, $^{24}$Mg, $^{25}$Mg, $^{26}$Mg, and $^{27}$Al, respectively, and the isotopic signals were measured on flat-top ion beam peaks at a mass resolving power of $\sim$6,000. The instrumental fractionation was corrected by using sample-standard bracketing. Samples and  isotopic standards were measured in alternating blocks of 20 cycles with integration times of $\sim$8s per cycle. Cross contamination between samples and standards was eliminated by rinsing the spray chamber with 2\% HNO$_3$ for 200 s after each block. The  signals of $^{56}$Fe and $^{24}$Mg in the rinse were $\sim$10mV and $\sim$8 mV, respectively. Sample and standard solutions were diluted to $\sim$2-4 ppm in 2\% HNO3 for isotopic analyses. The sensitivity of  $^{56}$Fe and $^{24}$Mg was typically 2 V/ppm and 7 V/ppm (1$\times10^{-10}$ amps/ppm). Uncertainties for each experimental datum are reported as 2 standard errors (2se) obtained from the mass spectrometer measurements, representing internal precision.  Uncertainties in the USGS standards are 2se (uncertainties in the mean) from multiple measurements ($n=9$ for Fe in BHVO-2, $n=6$ for Mg from BHVO-2, and $n=3$ for DTS-02), and represent external precision. We used IRMM-14 and DSM-3 as the standards for bracketing and as the primary isotopic standards for reporting our Fe and Mg isotope ratios.

The chemical compositions of our sample solutions were also performed on the MC-ICPMS at UCLA. Prior to purification, we set aside $\sim$10-20\% of each of the digested sample solutions to provide samples free of  chemical purification.  These samples were evaporated and redissolved in 2\% HNO$_3$ for MC-ICPMS analyses of Fe/Al and Mg/Al ratios. Ion beams for $^{24}$Mg and $^{56}$Fe were obtained in the center faraday cup by peak jumping as necessary, and compared with the ion beam for Al. In order to resolve  interferences from $^{40}$Ar$^{14}$N$^{+}$ and $^{40}$Ar$^{16}$O$^{+}$ we used a mass resolving power of  $>$8500 as was done for the isotope ratio measurements. Samples were measured in alternating blocks of 5 cycles with integration times of $\sim$8s per cycle, and the washout time was 150 s after each block.

\section*{Appendix 2: Elemental fractionation and activity ratios}

In order to characterize the fractional losses of Fe and Mg in our experiments, we used our measured Fe/Al and Mg/Al ratios.  Because Al is highly refractory, to a good approximation the total mass of Al is preserved in the melt during the experiments.  The ratio Fe/Al relative to the initial ratio is therefore a measure of Fe/Fe$_{\rm o}$ where Fe is the mass (or moles) of Fe after evaporation and Fe$_{\rm o}$ is the initial mass (or moles) of Fe prior to evaporation. To see this, the Fe/Al ratios can be expanded as follows:

\begin{equation}
\frac{ [\mathrm{Fe}]/[\mathrm{Al}]}
{[\mathrm{Fe}]_{\rm o}/[\mathrm{Al}]_{\rm o]}}=
\frac{ \mathrm{Fe}}{ \mathrm{Fe}_{\rm o}}   \frac{ \mathrm{Al}_{\rm o}}{ \mathrm{Al}}  =
\frac{ \mathrm{Fe}}{ \mathrm{Fe}_{\rm o}}  
\label{Eqn:masstrick}
\end{equation}
where [Fe] ([Al]) represents the concentration of Fe (Al) by mass and $ \mathrm{Fe}/ \mathrm{Fe}_{\rm o}$  ($ \mathrm{Al}/ \mathrm{Al}_{\rm o}$) is the mass or molar ratio of Fe (Al) relative to the initial value.  Here we make use of the fact that $\mathrm{Al}_{\rm o}/ \mathrm{Al}$ is unity.  We emphasize that it is the mass of Al that is constant, an extensive quantity, not the intensive concentrations.  Similar equations apply for Mg.  These Al-normalized ratios have the advantage of not being affected by closure, unlike concentrations (e.g., weight fractions).  Closure causes spurious correlations among variables \citep{Chayes1971}, as will become important in the analysis below.

We define $\mathrm{Fe}/ \mathrm{Fe}_{\rm o}$ and $\mathrm{Mg}/ \mathrm{Mg_{\rm o}}$ as the fraction of Fe and Mg remaining, $f_{\rm Fe}$ and $f_{\rm Mg}$, respectively. The fractional losses for Fe and Mg are related to the net evaporative fluxes of the elements using 

\begin{equation}
\frac{
(f_{\rm Fe}-1)}
{(f_{\rm Mg}-1)}=
\frac{ F_{\rm Fe, net}}
{F_{\rm Mg, net}}
\frac{x_{\rm Mg}^{\rm o}}{x_{\rm Fe}^{\rm o}},
\label{Eqn:fluxratio}
\end{equation}\

\noindent where $x_i^{\rm o}$ is the mole fraction of species $i$ for the melt prior to evaporation and $F_{i,{\rm net}}$ is the net evaporative flux of element $i$.

We can use our net evaporation losses of Fe and Mg to constrain the thermodynamic behavior of these elements in these melts of enstatite chondrite composition. To see this, we present the equations for evaporative fluxes in general.  The net evaporative flux of element $i$, in our case Fe and Mg, can be described as a balance between the free evaporation flux that is proportional to the equilibrium vapor pressure of $i$, $P_{i,{\rm sat}}$, and the return flux that is proportional to the ambient partial pressure $P_i$ by the Hertz-Knudsen equation

\begin{align}
F_{i,{\rm net}} =& 
F_{i,{\rm evap}}-F_{i,{\rm return}} \nonumber \\
=&\frac{1}{   \sqrt{2\pi m_i R}  } \left( \gamma_e 
\frac{  P_{i,{\rm sat}}  }{  \sqrt{T_{\rm melt}}  } - 
\gamma_c \frac{P_i}{   \sqrt{T_{\rm gas}}  } \right),
\label{Eqn:fullHK}
\end{align}\

\noindent where $R$ is gas constant, $T$ is temperature, $m_i$ the molecular mass of $i$ per mole, and $\gamma_e$ and $\gamma_e$ are the empirical evaporation and condensation coefficients.  We will simplify Equation \ref{Eqn:fullHK} by assuming that the melt and gas at the surface of the melt have the same temperature and that the evaporation and condensation coefficients are unity for evaporation and condensation involving melts \citep{Sossi2019}.  Equation \ref{Eqn:fullHK} illustrates the utility of defining the saturation for the system, S, where $\mathrm{S}=P_i/P_{i,{\rm sat}}$, $P_i = x_i P$ is the partial pressure of species $i$, $x_i$ is the mole fraction of species $i$,  $P$ is total pressure, and $P_{i,{\rm sat}}$ is the equilibrium partial pressure at saturation.  Where S$_i$ is unity, there is no net flux and the system is at equilibirum, while S$_i < 1$ results in evaporative losses.  With this definition for saturation, it is straightforward to show that \citep{Young2019}

\begin{equation}
F_{i,{\rm net}}=F_{i,{\rm evap}} (1-\mathrm{S}_i).
\label{Eqn:fvsS}
\end{equation}\

The  saturation vapor pressures of Fe and Mg can be obtained from thermodynamics by envisioning that the species in the melt that liberate Fe and Mg are the simple oxides FeO and MgO.  The resulting expression for free evaporation is

\begin{equation}
F_{\rm M, evap}=a_{\rm MO}^{\rm melt}
\frac{\exp({-\Delta\hat{G}_{\rm rxn}^{\rm o}/(RT)})}{\sqrt{2\pi RT m_{\rm M}}}  P_{\rm O_2}^{-1/2}
\label{Eqn:thermo}
\end{equation}

\noindent where $-\Delta\hat{G}_{\rm rxn}^{\rm o}/(RT)$ is the logarithm of the equilibrium constant for the reaction MO $\rightleftharpoons \mathrm{M} + 1/2 \mathrm{O}_2$ for metal M (either Fe or Mg in our case) expressed in terms of the standard state Gibbs free energy for the reaction.  Equations \ref{Eqn:thermo}, \ref{Eqn:fvsS}, and \ref{Eqn:fluxratio} 
can be combined to give an expression relating the fractional losses of Fe and Mg in our experiments to standard thermodynamic parameters:

\begin{equation}
\frac{f_{\rm Fe}-1}{f_{\rm Mg}-1}=\frac{x_{\rm Mg}^{\rm o}}{x_{\rm Fe}^{\rm o}}
\frac{\gamma_{\rm Fe}x_{\rm Fe}}{\gamma_{\rm Mg}x_{\rm Mg}}
 \exp{
 \left( \frac{\Delta\hat{G}_{\rm Mg,rxn}^{\rm o}-\Delta\hat{G}_{\rm Fe,rxn}^{\rm o}
 }
 {RT}\right)
 }
\sqrt{\frac{m_{\rm Mg}}{m_{\rm Fe}}}
\frac{(1-\mathrm{S}_{\rm Fe})}{(1-\mathrm{S}_{\rm Mg})}.
\label{Eqn:totalratio}
\end{equation}\

\noindent
Here we have expanded the activities in the melt in terms of mole fractions $x_i$ and activity coefficients $\gamma_i$ (not to be confused with the evaporation or condensation coefficients that we have assumed are unity for now).  For the beginning stages of evaporation, where composition has changed little, the mole fractions cancel, and the only unknown parameters in Equation \ref{Eqn:totalratio} are the activity coefficients and the saturation for each element.  We showed  in the discussion of isotope fractionation that the flow of gas over the molten spheres is the primary control on saturation, and that $\mathrm{S}_{\rm Fe}$ and $\mathrm{S}_{\rm Mg}$ are indistinguishable in these experiments so the corrections for saturation cancel.  Therefore, the slope of a line fitting the data for $f_{\rm Fe, net}-1$ vs. $f_{\rm Fe, net}-1$ can be used to constrain the ratio of activity coefficients in the melts. 

\begin{figure}
\centering
\captionsetup{width=0.95\linewidth}
 \includegraphics[width=4.0in]{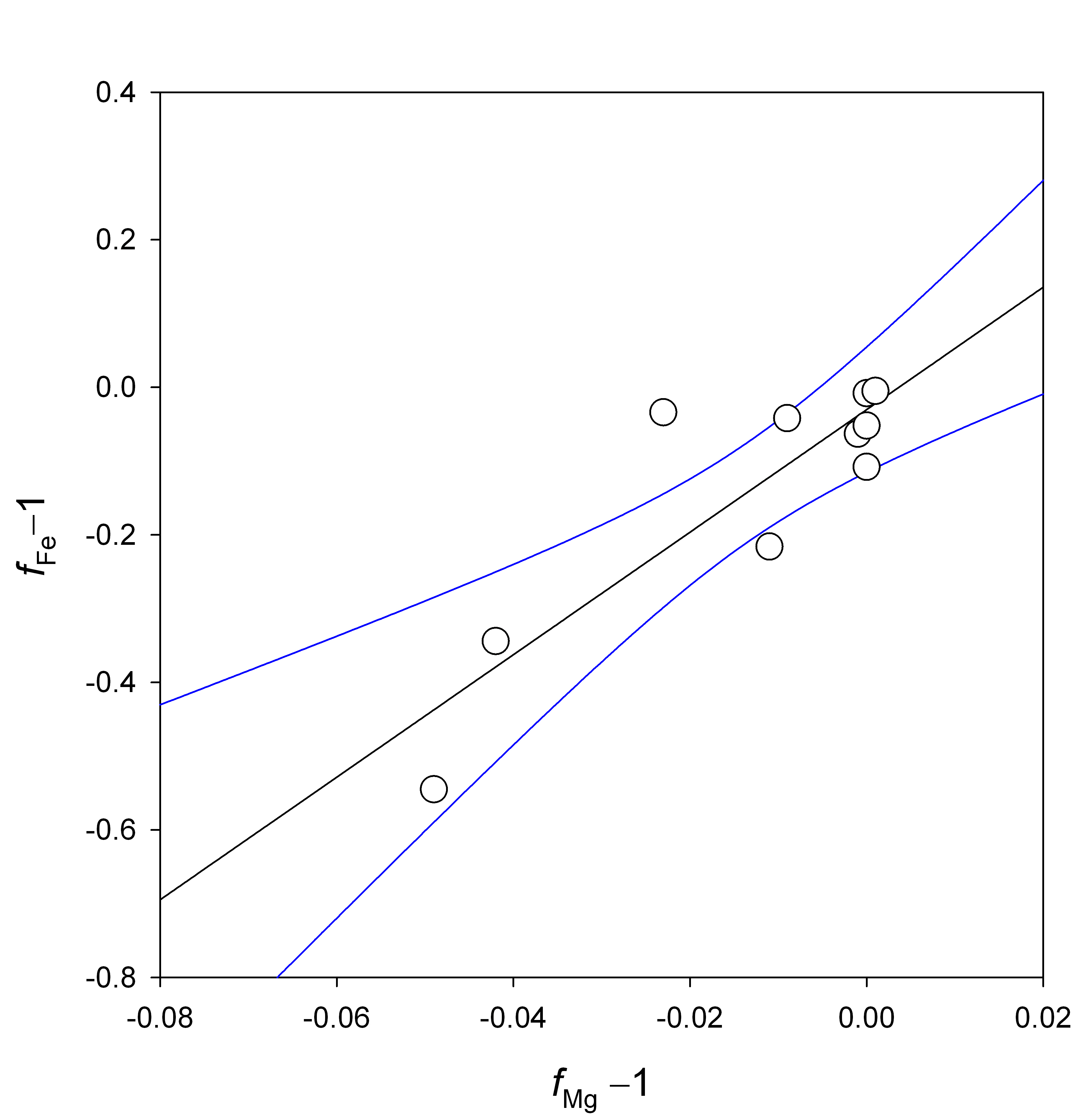}
\caption{Plot of correlation between fractional losses of Fe and Mg as defined in Equation \ref{Eqn:fluxratio}.  The parameters $f_{\rm Mg}$ and $f_{\rm Fe}$ are the fractional amounts of Mg and Fe remaining in the melts, as defined in the text.  The slope of the linear fit is $8.3\pm 3.3$ and defines the activity coefficient ratio $\gamma_{\rm Fe}/\gamma_{\rm Mg}$ as described in the text. 95\% error envelope for the fit is shown for  reference.  Only the data for N$_2$ gas are used for consistency. } 
\label{Fig:elementresults} 
\end{figure}

The linear best-fit for the Fe and Mg concentrations for our N$_2$ experiments is $8.3\pm 3.3$.  The predicted slope from Equation \ref{Eqn:totalratio} based on the temperature-dependent  thermochemical data from the NIST database, and the assumption that the activity coefficient ratio is unity, is $0.48$. Matching the slope using Equation \ref{Eqn:totalratio} therefore requires that the ratio $\gamma_{\rm Fe}/\gamma_{\rm Mg}$ for these melts instead is significantly greater than 1. Given the uncertainty in the slope, the $2\sigma$ range in $\gamma_{\rm Fe}/\gamma_{\rm Mg}$ is from $12$ to $24$.  

We can compare our derived activity coefficient ratio with the ratio returned by the MAGMA code \citep{Fegley1987,Schaefer2009}.  In this code,  widely used in planetary science and astrophysics, evaporation is  cast in terms of reactions involving the simple oxides (MgO, SiO$_2$, FeO, Na$_2$O) but allowing for the fact that the activities of these components are known to be quite low compared with their concentrations in the liquid.  Recalling that activity of a component $i$ is the effective concentration given by $a_i = \gamma_i x_i$ where $x_i$ is mole fraction of the component in the melt, and $\gamma_i$ is the activity coefficient, there are two ways to account for this low activity relative to mole fractions.  One is to assign a low value to the activity coefficient $\gamma_i$, but these are in many cases not well known based on limited experiments. The other is to consider that the way that Fe and Mg exist in a silicate liquid is not mainly as FeO and MgO, but rather as FeSiO$_3$, MgSiO$_3$, and other similar silicate mineral-like species through reactions like MgO + SiO$_2$ = MgSiO$_3$ in the melt.  The activities of the simple oxides can be accounted for to first order by such reactions, as described by \cite{Hastie1986}.  This is the approach used in the MAGMA code. For the bulk composition of our synthetic enstatite chondrite melts at the temperature of the experiments, the MAGMA code returns $\gamma_{\rm Fe}$ and $\gamma_{\rm Mg}$ values of $0.789$ and $0.135$, respectivley, and the ratio is $5.84$.  Our measured ratio is 2 to 5 times this value, providing an experimental constraint on activities of FeO and MgO in these melts.   

Equation \ref{Eqn:totalratio} applies where evaporation is independent of evolving concentrations, as in the case of congruent evaporation of major elements.  Another approach was taken by \cite{Sossi2019} where these authors assume the evaporating elements behave as trace elements, resulting in replacement of the ratio $(f_{\rm Fe}-1)/(f_{\rm Mg}-1)$ on the left-hand side of Equation \ref{Eqn:totalratio} with $\ln{f_{\rm Fe}}/\ln{f_{\rm Mg}}$.  If applied to our data, the activity coefficient ratio increases from 12 to 24, to  14 to 33.

\section*{Appendix 3: Gas temperature near the evaporating sphere}

We can evaluate the prospects for heating the levitation gas near the evaporating sphere using the thermal transport P\'{e}clet number that quantifies the ratio of the rate of advective transport to the rate of thermal diffusive transport, where ${\rm Pe} =V_d L/\kappa$, $\kappa$ is the thermal diffusivity and $L$ is the length scale of interest.  For a thermal diffusivity in N$_2$ gas of $2\times 10^{-5}$ m$^2$/s at ambient $T$ \citep{Salazar1999} and $L$ equal to the diameter of the evaporating sphere, ${\rm Pe} = 4100$.  The value for Pe approaches unity only for  $L < 1\mu {\rm m}$, indicating the dominance of velocity over thermal diffusion in determining temperature more than a micron from the sphere.  This is consistent with estimates for the width of the thermal boundary layer afforded by heat flux balance in the boundary.  The flux balance is $\vec{u} \rho  C T_{\rm ambient} + k(-\nabla T)_{\vec{n}} =\vec{u} \rho  C T_{\rm out}$ where the gradient refers to the direction normal to the surface, $\vec{u}$ refers to the flow across the surface, $C$ is the heat capacity of the gas, $k$ is thermal conductivity, $T_{\rm ambient}$ is the ambient temperature of the gas and $T_{\rm out}$ is the temperature of the gas heated by the molten sphere.  Rearranging, we have $-(\nabla T )_{\vec{n}}=(\vec{u}/\kappa)\Delta T$ and the width of the thermal boundary layer is then $\Delta T/(-\nabla T )_{\vec{n}}=\kappa/\vec{u}$.  For $\vec{u} \sim V_d = 41$m/s, the thermal boundary layer is estimated to be $0.5 \mu {\rm m}$.  For gases in general, the width of the thermal boundary layer should be similar to the width of the velocity boundary layer (Prandtl number $\sim 1$) where the flow is laminar (e.g., where $\theta > \pi/2$, Figure \ref{Fig:schematic}), implying that where the stream velocity applies, the ambient gas temperature prevails.  

\end{document}